# Glyme-based electrolytes: suitable solutions for next-generation lithium batteries

Daniele Di Lecce[1], Vittorio Marangon[1,2], Hun-Gi Jung[3], Yoichi Tominaga[4,5], Steve Greenbaum[6], Jusef Hassoun[1,2,5,7,*]

[1] *Graphene Labs, Istituto Italiano di Tecnologia, via Morego 30, Genova, 16163, Italy*

[2] *University of Ferrara, Department of Chemical, Pharmaceutical and Agricultural Sciences, Via Fossato di Mortara 17, 44121, Ferrara, Italy.*

[3] *Center for Energy Storage Research, Korea Institute of Science and Technology (KIST), Hwarangno 14-gil 5, Seongbuk-gu, Seoul, 02792, Republic of Korea*

[4] *Tokyo University of Agriculture and Technology, Graduate School of Bio-Applications and Systems Engineering (BASE), 2-24-16, Naka-cho, Koganei-shi, Tokyo 184-8588, Japan.*

[5] *Institute of Global Innovation Research (GIR), Tokyo University of Agriculture and Technology, Tokyo, Japan*

[6] *Hunter College of CUNY, 695 Park Avenue, NY 10065, New York, USA*

[7] *National Interuniversity Consortium of Materials Science and Technology (INSTM) University of Ferrara Research Unit, University of Ferrara, Via Fossato di Mortara, 17, 44121, Ferrara, Italy.*

Corresponding Author: jusef.hassoun@iit.it, jusef.hassoun@unife.it

**Table of Contents**

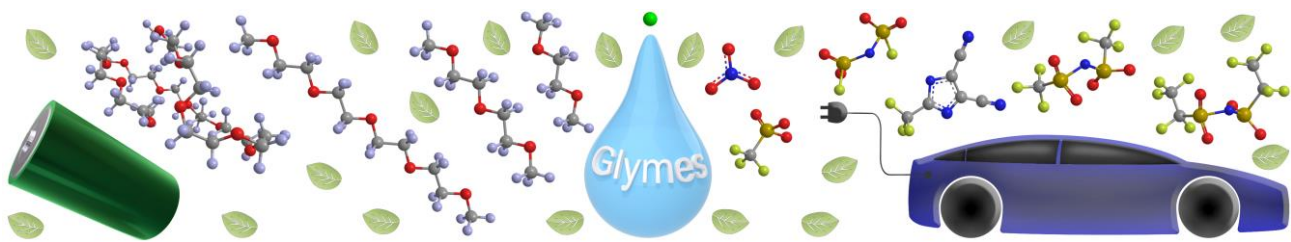

Glymes are possible electrolyte solvents for high-energy lithium batteries due to suitable thermal and electrochemical properties. The most relevant results gathered over twenty years of scientific research on glyme-based electrolytes are reviewed, and possible approaches to achieve enhanced lithium-metal cells using insertion or conversion cathodes are suggested and discussed.

**Keywords**

**Abstract**

The concept of *green* in a battery involves the chemical nature of electrodes and electrolytes as well as the economic sustainability of the cell. Although these aspects are typically discussed separately, they are deeply interconnected: indeed, a new electrolyte can allow the use of different cathodes with higher energy, lower cost or more pronounced environmental compatibility. In this respect, we focus on an alternative class of electrolyte solutions for lithium batteries formed by dissolving LiX salts in glyme solvents, i.e., organic ethers with the molecular formula $CH_3O[CH_2CH_2O]_nCH_3$ differing by chain length. The advantages of these electrolytes with respect to the state-of-the-art ones are initially illustrated in terms of flammability, stability, toxicity, environmental compatibility, cell performances and economic impact. A particular light is shed on the stability of these systems, particularly in the polymer state, and in various environments including oxygen, sulfur and high-energy lithium metal. Subsequently, the most relevant studies on the chemical-physical features, the characteristic structures, the favorable properties, and the electrochemical behavior of the glyme-based solutions are discussed, and the most recent technological achievements in terms of cell design and battery performance are described. In the final sections, the use of glyme-based electrolytes in high-energy cells arranged by coupling the lithium-metal anode with conventional insertion cathodes as well as in alternative and new batteries exploiting the Li-S and Li-O$_2$ conversion processes are described in detail. The various paragraphs actually reveal the bonuses, including safety, low cost and sustainability, that can be achieved by employing the glyme-based electrolytes with respect to the commercially available ones, in particular taking into account future and alternative applications. Particular relevance is given by the glymes with long chain that reveal a remarkable stability, high safety and very low toxicity. Therefore, this review is expected to shed light on the potentialities, the actual advantages compared to the state-of-the-art battery, and the possible applications of electrolytes based on glyme solvent in a next-generation of energy storage systems.

**Introduction**

Several literature papers proposed for improving the *sustainability* of the electrochemical energy storage systems focused on the *green nature* of the various cell components. On the other hand, the most significant approach for achieving the sustainability and the green concept foresees the demonstration of actual advances in comparison with the existing technologies.[1] In this respect, a widespread diffusion of lithium-ion (Li-ion) batteries over the past decades has been driving a notable technological change in our society and is currently enabling a gradual transition to a more environmentally sustainable automotive mobility.[2] Smartphone, laptops, and tablets powered by lithium-ion cells are among the most used devices in our daily life, and hybrid and full electric vehicles are becoming widely adopted, thanks to several programs lunched worldwide to decrease environmental pollution.[3] Furthermore, Li-ion batteries could play a key role in developing smart grids coupled with renewable energy sources,[4] as suggested by the already installed and operating battery packs for stationary storage.[5] Driven by the undeniable success of the Li-ion technology, ambitious targets have been set by several countries to further boost the cell performance by increasing the energy density, and the lithium-metal battery design has been identified as one of the most promising new systems.[6] Despite a few practical examples of commercial lithium-metal batteries, various challenges still have to be overcome to develop high-energy cells with a suitable safety level and a reliable long-term behavior.[6] In this regard, the characteristic properties of the most common electrolyte solutions for Li-ion batteries pose some issues.[7] These solutions typically consist of mixtures of a lithium salt (e.g., $LiPF_6$) and organic ester solvents, such as alkyl carbonates, and indeed exhibit high conductivity which enable a satisfactory battery performance in terms of capacity, energy density, and rate capability.[7] On the other hand, these electrolytes are volatile, flammable, and poorly stable in contact with lithium metal, in particular during prolonged operation.[8–10] Therefore, several research studies have been focusing since nineties on optimizing alternative electrolyte solutions,[11–15] such as those prepared by dissolving a lithium salt in a lowly volatile and modestly flammable ether oligomer with a –(CH$_2$CH$_2$O)– unit, that is, ethylene oxide (EO)-glymes.[16–24] EO-based glymes can be synthesized from ethylene epoxide by various large scale methods, including reaction in alcoholic environment with sodium following the Williamson mechanism, methylation of glycol ether with methyl sulfate, Lewis acid-catalyzed cleavage of ethylene oxide by ether, and reaction of ethylene glycol with alcohol catalyzed polyperfuorosulfonic acid resin at high temperature and pressure.[25] Instead, aliphatic carbonate presently used in the battery electrolyte formulation may be industrially achieved by more complex organic pathways including phosgenation, oxidative carbonylation, reaction of urea with alcohols, reaction of oxiranes with carbon dioxide, through metal carbonate, and by carbonate interchange reaction.[26] In particular, dialkyl carbonate such as DMC and DEC, may be prepared from halohydrins, and from alcohols and carbon monoxide by elemental sulfur, while alicyclic carbonate such as EC may be obtained from the corresponding halogenated carbonates.[26] For reader's convenience, Table 1 reports the acronyms of the several chemical species discussed herein.

**Table 1**

In addition to the simplest and more environmental friendly preparation pathway compared to the common electrolyte solvents used in battery (i.e., EC, DMC, EMC, DEC), the glymes, and particularly those with the longer chain length (G$_3$, G$_4$, PEGDME), are characterized by a more relevant safety content, lower flammability and flash point as well as less relevant toxicity except for possible issues on fertility expected for glymes with the lower chain length. Table 2 displays the physical-chemical properties of the conventional carbonate solvents and glymes used for battery application, including the related safety hazards as indicated by the SDS data (Sigma-Aldrich) and confirmed by literature.[25,27–31] PC and EC are the only carbonate solvents without flammability, however EC has an elevated toxicity. Instead, EC and DMC which are typically combined as the electrolyte solvents for LIBs are highly flammable, thus suggesting the need for a safer alternative.[32] On the other hand, glymes show a favorable trend associated to the lengthening of the -CH$_2$CH$_2$O-chains. Indeed, DME ($G_1$) and DEGDME ($G_2$) are the only glyme species to show relevant flammability, while TREGDME ($G_3$) and TEGDME ($G_4$) exhibit lowly toxicity and liquid PEGDMEs (250 and 500 g mol$^{-1}$) are classified as non-dangerous for humans. Despite DME is usually employed in Li-S battery due to the enhanced stability towards the lithium polysulfide intermediates formed by the conversion electrochemical process, the safety issues related to its relevant volatility and flammability are by now acknowledged, and the search for safer electrolyte solutions to achieve Li-S devices of practical interest is a deeply investigated topic.[33] Moreover, decrease of the electrolyte flashpoint may be achieved by using non-flammable co-solvents and flame retardant additives, such as phosphorous-containing (e.g., TMP, TEP, TPrP, TFEP) and fluorinated (e.g., FEMC, FEC, DFDEC, TTFE, PFPN) species,[34] or by functionalization of the glyme solvent.[35] Among the various strategies, solid electrolyte configurations may represent a viable strategy to increase the safety content of the battery due to enhanced chemical, thermal and mechanical stability. Ceramic electrolytes such as garnet-type $Li_7La_3Zr_2O_{12}$ (LLZO) or NASICON-derived structures, e.g., $Li_{1.5}Al_{0.5}Ge_{1.5}(PO_4)_3$ (LAGP), provide fast Li$^+$ transport and suitable ionic conductivity at room temperature, despite the cycling behavior may be affected by the high interphase stability due to a poor contact between electrodes and electrolyte in the cell.[36] On the other hand, solid polymer electrolytes benefit from suitable electrode/electrolyte contact and remarkable conductivity which, however, are reached at medium-high operative temperatures. Indeed, PEO-based electrolytes usually require temperatures above 65 °C to allow proper amorphization of the crystalline structure and satisfactory battery performance,[37] which may be improved through the introduction of copolymer blocks such as PS, PEGMA, PEGDMA or PEGA,[36,38] or ceramic fillers such as $SiO_2$, $ZrO_2$ or $TiO_2$ in the PEO matrix.[37,39,40] The substitution of PEO with polycarbonate species was recently considered due to the lower crystallinity and good oxidative stability, even though the low stability towards lithium metal may limit their application.[36] Furthermore, alternative polymer chemistries can be employed to stabilize the lithium anode and synthetize effective separators in order to enhance the Li$^+$ exchange between the electrodes.[41,42]

Table 2

Along with the initial conceptualization and pioneering studies of glyme-based electrolytes, Li-metal batteries using the most common intercalation electrodes, such as $LiCoO_2$ and graphite, as well as other insertion cathodes such as $LiFePO_4$, were proposed with promising results, in spite of several issues which were firstly identified.[43–60] Afterwards, "high-concentration" glyme-based electrolytes attracted a great deal of attention due to their favorable properties, and an intriguing dilemma on the actual nature of these mixtures, which have been described as either solvent-in-salt solutions or solvated ionic liquids (SILs), was posed.[61–82] Moreover, glyme-based electrolytes have been gaining renewed interest due to a possible suitability for the emerging high-energy lithium-sulfur (Li-S) battery, which is formed by combining a lithium-metal anode and a sulfur-based cathode.[83–102] This new technology is nowadays considered close to a practical application and holds the promise of a breakthrough in storable energy per unit mass.[103,104] Furthermore, the glyme-based solutions have been selected as the electrolytes of choice for the lithium-oxygen (Li-$O_2$) cell, in which a lithium-metal anode is coupled with a gas diffusion layer electrode enabling the $O_2$ electrochemical conversion thanks to an open design. Notably, this system could store even more energy than the Li-S cell and has been suggested as possible battery for future applications.[105–130] In this review, we discuss with chronological detail these various developments in the research on glyme-based electrolytes for lithium batteries, which are summarized by the scheme in Figure 1 (panel a). For the reader's convenience the various techniques and expressions cited throughout the following sections are listed and explained in Tables 3. Furthermore, the panel b of Fig. 1 reports a comparison between the present Li-ion battery and the two emerging energy storage systems (i.e., Li-S and Li-$O_2$ cells), which can be allowed by the use of glyme-based electrolyte for possible application in the electric vehicles. The figure suitably reveals the advances potentially achieved by these new system both in terms of driving range (km by a single charge),[131] and in terms of the economic impact (USD per kWh of battery pack).[132]

Table 3, Figure 1

## 1. The "glyme electrolyte": initial studies and fundamental concepts

Pioneering studies on possible energy-storage applications of glymes were carried out in the late 1990s, mostly driven by encouraging results on the lithium-metal polymer battery employing PEO.[6] Indeed, electrolytes using low-molecular-weight glymes exhibited compromise characteristics in terms of chemical stability and viscosity, thereby holding the promise of an enhanced safety level as compared to that of conventional alkyl-carbonate-based solutions, along with suitable Li$^+$ transport properties at room temperature.[19,53,86] On the other hand, understanding the complex Li-electrode/electrolyte interface in glyme-based cells has required several studies, which are still ongoing to fully clarify the characteristic features of such an intriguing and intrinsically safe system. Initial works opened a debate on the actual suitability of this class of electrolyte solutions in high-energy batteries owing to doubtful results on the compatibility of glyme solvents with lithium metal. Accordingly, a first comparative investigation of the Li electrode in solutions of various salts, that is, LiAsF$_6$, LiClO$_4$, LiTf, LiTFSI, LiBF$_4$, LiBr and LiI, in DME, EG, and G$_1$ solvent, with DOL, PC, EC and DMC as co-solvents, suggested poor surface chemistry and rough morphology of the metal electrode after electrodeposition-dissolution, which lead to a low cycling efficiency in rechargeable batteries. In this regard, cyclic voltammetry provided insights on the surface films deposited in the cell, as shown in Figure 2a, which reports a steady-state profile of a polycrystalline gold electrode in solutions using the above-mentioned solvents and LiTf as salt. The Li underpotential deposition (UPD) and stripping are marked in the figure along with the non-Faradaic region.[16] However, other studies demonstrated promising characteristics of glyme-based solutions of LiTFSI (G$_n$ with *n* from 1 to 4), revealing for this salt high compatibility with various solvents and high electrochemical stability. Notably, LiTFSI appeared strongly associated in glymes and moderately associated in TESA at low concentrations, formed stable solvates in G$_1$ at intermediate concentrations, and displayed thermodynamic properties approaching those of the molten salts at high concentrations. Figure 2b shows the trend of specific conductivity as a function of the concentration of LiTFSI in $G_n$ solvents. Reasonably high specific conductivity was observed for both LiTFSI and LiClO$_4$ in various solvents, which indicated that the ionic conductance at high concentration in solvents of low dielectric constant was limited by a charge transfer process rather than by the migration of free ions.[17] To verify the hypothesis of stable solvates that persist in the solution influencing its properties, phase diagrams and Raman spectra have been measured for mixtures of LiTFSI and ACN, PC or glymes ($G_n$, with $n = 1$, 4, and 10). In fact, the systems without solvates show relative intensities of the solvent and salt Raman bands which are proportional to the concentration. On the other hand, important changes in the relative intensities of these band reflect the presence of stable solvates in the electrolyte, which are additionally detected in the phase diagrams. Moreover, X-ray crystallography reveals free ions, SSIPs and CIPs in the solutions, suggesting that no stable solvates are formed in $(G_n)_x$:LiTFSI electrolytes for $n > 2$.[18] The interplay between lithium-ion solvation state, ionic conductivity, and charge-discharge cycling efficiency of Li-metal anode was further investigated. In particular, ternary mixed solvent consisting of $G_n$ ($n$ from 1 to 4), EC, and EMC, dissolving LiPF$_6$ (1M) were comparatively studied. These glyme solutions exhibited higher conductivity and higher lithium cycling efficiency than EC/EMC, while both the conductivity and the viscosity typically increased as the ethylene-oxide chain length decreases (decrease in $n$). Notably, this decrease in viscosity was associated with a change in the lithium-ion solvation structure, which occurred when a glyme was added to EC/EMC and was caused by a selective solvation of the glyme with respect to lithium ions as demonstrated by $^{13}$C-NMR measurements. The lithium cycling efficiency value depended on the charge-discharge current ($I_{ps}$), and the ethylene-oxide number, $n$, affected this trend. In particular, when $n$ increased there was a decrease in the $I_{ps}$ exhibiting the maximum value of efficiency (Eff$_{max}$, see Figure 2c). A similar change in conductivity with $n$ was observed (see Figure 2c). Among the various glymes taken into account, $G_2$ or $G_3$ exhibited very promising characteristics, namely high conductivity and suitable charge-discharge cycling behavior at a high current.[19] In a subsequent report,[20] the crystal structures of glyme solvates with LiTFSI, and LiBETI were determined, and order-disorder solid phase transitions in many of the solvates were identified. Further work was carried out to shed light on the molecular interactions in glyme-based solutions. Accordingly, phase diagrams of mixtures of $G_n$ ($n$ = 1, 2, 3, and 4) and LiBETI, LiAsF$_6$, LiI, LiIO$_4$, LiBF$_4$, LiTf, LiBr, LiNO$_3$, and LiCF$_3$CO$_2$ were proposed, the relations between ionic association strength of the salt (that is, anion characteristics), chain length of the glyme, solvate formation, and thermophysical properties of the mixture were investigated, thereby providing a comprehensive model of solvate formation and ionic interactions in these electrolyte systems. Figure 2d illustrates an approximate ordering for increasing ionic association strength of LiX salts in glymes.[21] Glymes were suggested to form solid complexes with lithium salts, which would be suitable as "soft" solid electrolytes for Li batteries exhibiting a wide range of ion transport properties. Thus, complexes between LiAsF$_6$ and $G_n$ ($n$ = 3 and 4) showed significantly different cation transference number, resulting from the presence of channels for Li$^+$ migration in G$_3$ and from weak binding of the AsF$_6^-$ in the structure of G$_4$.[22] Considerable efforts were devoted to understanding the ion-solvent arrangements in the glyme-based electrolyte solutions. For instance, $^{19}$F NMR spectroscopy and conductivity data were collected to determinate the ion pair formation constants of LiTFB and LiTf solutions in mixtures of DOL and G$_1$, G$_2$, or water, and the obtained results were interestingly similar to those of liquid PEGDME and solid PEO.[23] Furthermore, the formation constants of ionic pairs solution of lithium salts in G$_1$, G$_2$, and G$_3$ were estimated using $^7$Li, $^{11}$B and $^{19}$F NMR analyses, which demonstrated that even in solvents with very similar coordination (in terms of the donor and acceptor number values) and dielectric properties, the ionic pair formation constant depended on effects related to ion agglomerate formation, non-covalent interactions between ions and liquid matrix, as well as the number of interacting centers in the solvent molecules.[24]

Therefore, glymes are proposed for LIBs consequently to the promising features of PEO-based electrolytes, with a stability depending on the specific glyme-salt (G-S) combination. In particular, the salt concentration appears to remarkably influence the electrolyte characteristics in terms of solvation ability, dissociation degree, and ionic conductivity. The formation of stable G-S *solvates* is hypothesized and preliminarily verified for specific compositions, while it is excluded for others. Furthermore, the Li/electrolyte interphase characteristics and the charge/discharge efficiency apparently depend on the adopted G-S combination, due to the effects of the electrolyte viscosity and the structure of the ion-solvent species. Likely, the better efficiency is ascribed to electrolytes formed by using glymes with increased chain length (i.e., with increasing n in $G_n$). In addition, phases with order-disorder structure are observed in some S-G solvates and suggested to affect the ion association degree of the salts, thus modifying the performance of the electrolyte in lithium cell.

**Figure 2**

## 2. Li-metal batteries using "glyme electrolytes" and insertion/intercalation electrodes

Glyme-based electrolytes have shown promising performances in lithium-metal batteries using insertion cathodes. A Li|LiCoO$_2$ call employing the [Li(G$_4$)][CTFSI] complex was assembled and tested, revealing a stable galvanostatic response during 50 cycles. LiCTFSI may form solid and liquid complexes with G$_3$ and G$_4$, respectively. Notably, the electrolyte was stable in the wide potential range from 0 to 4.5 V *vs* Li/Li$^+$ and had a much higher thermal stability as compared to that of pure G$_4$, whilst the vapor pressure of [Li(G$_4$)][CTFSI] was negligible at temperatures lower than 100 °C. Furthermore, the [Li(G$_4$)][CTFSI] complex exhibited an ionic conductivity of 0.8 mS cm$^{-1}$ at 30 °C, which was slightly lower than that of conventional alkyl-carbonate-based electrolyte solutions and higher than that of the [Li(G$_3$)][CTFSI] complex (see Figure 3a), in spite of a relatively high viscosity due to the high molar concentration (ca. 3 mol dm$^{-3}$). The pulsed gradient spin-echo NMR (PGSE-NMR) method was employed to measure the self-diffusion coefficients of Li$^+$ cation, CTFSI$^-$ anion, and glyme molecule, and the ionicity (dissociativity) of [Li(G$_4$)][CTFSI] at 30 °C was estimated to be ca. 0.5.[43] Another example showed that increasing the amount of glyme in G$_4$–LiTFSI complexes (where the molar ratio of G$_4$ ranged from 40 mol% to 60 mol%) decreases the viscosity and increases the ionic conductivity, thereby improving the rate capability of Li|LiCoO$_2$ cells. In addition,

Li|graphite cells with SEI forming additives, such as VC, VEC, and 13PS, showed a cycling performance comparable to that of conventional carbonate-based electrolytes. Lithium cells using $Li_4Ti_5O_{12}$ and $LiFePO_4$ along with the $[Li(G_4)][TFSI]$ complex exhibited excellent reversibility due to an optimal working voltage range.[44] Besides, an electrolyte formed by $G_3$ and LiFSI in 1:1 molar ratio showed relatively high thermal stability as well as stable cycling using $LiFePO_4$ and graphite electrodes, with 82% of capacity retention at 100 cycles.[45]

A literature work suggested that the oxidative stability of glyme molecules can be enhanced by the complex formation with alkali metal cations, such as $[Li(G_n)_1][TFSI]$ with $n = 3$ and 4, which were classified as a room-temperature SIL consisting of a $[Li(G_n)_1]^+$ complex cation and a $TFSI^-$ anion due to their liquid state maintained over a wide temperature range and a high self-dissociativity (ionicity) at room temperature. The increase in salt concentration in the $[Li(G_n)_1][TFSI]$ equimolar SIL remarkably enhanced the oxidative stability of the solution to a potential as high as 5 V $vs$ $Li^+$/Li, likely owing to the donation of lone pairs of ether oxygen atoms to the $Li^+$ cation, which resulted in the lowering of the HOMO of the glyme as suggested by *ab initio* molecular orbital calculations. On the other hand, $[Li(G_n)_x][TFSI]$ solutions with $x > 1$ were stable up to 4 V $vs$ $Li^+/Li^+$ (see Figure 3b and c). NMR data indicated $Li^+$ transport in the equimolar complex via migration of the $[Li(G_3)_1]^+$ solvate, although the ligand exchange mechanism occurred as a result of the electrochemical reaction at the electrode/electrolyte interface in the $Li|[Li(G_3)_1][TFSI]|LiCoO_2$ cell. This battery exhibited a steady behavior for more than 200 charge-discharge with a voltage range of 3.0 – 4.2 V.[46] Further studies of the $Li|LiCoO_2$ cell using molten $Li(G_n)_1[TFSI]$ equimolar complexes were carried out to elucidate the relations between the $Li^+$ limiting current density under one-dimensional finite-diffusion conditions and the rate capability. Voltage drops and incomplete discharge of the cell were observed when the applied current was higher than the limiting current density, which may reflect a depletion of the lithium salt in proximity of the cathode or saturation at the anode interphase. Comparative tests of single-particle $LiCoO_2$ and an electrode sheet in a typical electrolyte solution based on PC, in a binary LiTFSI-(DEME)TFSI ionic liquid, and in a $[Li(G_n)_1][TFSI]$ molten complex suggested that the Li$^+$ transport in the solution controls the rate capability of the cells using the latter electrode.[47] Notably, the [Li(G$_3$)][TFSI] complex demonstrated a very promising performance in lithium-metal cells using LiFePO$_4$ and LiNi$_{1/3}$Mn$_{1/3}$Co$_{1/3}$O$_2$ cathodes. The battery using the former positive electrode operated at about 3 V (Figure 3d), delivering a rather stable capacity over 600 charge-discharge cycles (Figure 3g), whilst that using the latter had a working voltage of about 4 V (Figure 3e) and exhibited a cycling trend strongly depending on the upper voltage cutoff. Accordingly, the cell charged up to 4.2 V showed a capacity retention approaching 60% after 400 cycles. Further tests revealed the formation of a favorable Li/electrolyte interphase, thereby suggesting the possible applicability of this electrolyte formulation in lithium batteries.[48] In this regard, EQCM-A was effectively employed to measure the product of the viscosity (ηL) and density (ρL) of the electrolyte near the electrode surface, i.e., ηLρL, along with changes in mass. The collected data showed a decrease in the ηLρL value during lithium deposition and a sharp increase during lithium dissolution, which were ascribed to changes in concentration and dissolved state of Li$^+$ in proximity to the electrode. The latter increase with the Li dissolution may adversely affect the cation transference number, leading to a decrease in the anodic current in the battery.[49] Highly concentrated electrolytes have been then extensively investigated due to their suitable properties for use in lithium metal cells with insertion cathodes. A comparative study of molten mixtures of LiTFSI and various ether solvents (THF, G$_1$, G$_2$ and G$_3$), where the ratio of ether-oxygen atoms to Li$^+$ (i.e., O/Li) was fixed at four, revealed that the capacity of the Li|LiCoO$_2$ cell with [Li(THF)$_4$][TFSI] dramatically decreased during cycling, whereas a similar cell employing [Li(G$_3$)$_1$][TFSI] displayed a stable behavior with a coulombic efficiency higher than 99% over 100 cycles. These results were related to the oxidative decomposition of the solvents as well as to a persistent Al corrosion occurring in [Li(THF)$_4$][TFSI] and [Li(G$_1$)$_2$][TFSI], which contain shorter ethers, whilst the use of [Li(G$_3$)$_1$][TFSI] ensured effective suppression of the side reactions.[50] Literature also suggested that the ionic conductivity of glyme–LiTFSI solvate ionic liquids (SILs) may be inversely proportional to the viscosity, in agreement with the Walden's rule. Thus, a decrease in ionic conductivity with increasing concentration of LiTFSI was observed and associated with the formation of a bulky lithium species, i.e., [Li(TFSI)$_2$]$^-$.[51] The [Li(G$_4$)][TFSI] complex was employed in a quasi-solid-state electrolyte with fumed silica nanoparticles. This electrolyte was used in double-layered and triple-layered high-voltage bipolar stacked batteries, which showed a working voltage of 6.7 and 10.0 V, respectively, that is, two and three times that of the single-layered device (3.4 V). The double-layered device showed a capacity retention of 99% after 200 cycles at C/2.[52]

The characteristic features of the positive electrodes have crucial effects on the cell performance as they may influence the working voltage, the properties of the electrode/electrolyte interphase, and the rate capability. Accordingly, an electrolyte formed by dissolving 1 mol LiTf and 1 mol LiNO$_3$ in 1 kg of G$_4$ exhibited different behaviors in lithium-metal cells employing LiFePO$_4$ and LiMn$_{0.5}$Fe$_{0.5}$PO$_4$ cathodes. The cell using LiFePO$_4$ operated at 3.5 V with capacity ranging from 150 mAh g$^{-1}$ at C/10 to 110 mAh g$^{-1}$ at 2C, while the one employing LiMn$_{0.5}$Fe$_{0.5}$PO$_4$ showed two plateaus at 4.1 V and 3.5 V (Figure 3f) with capacity ranging from 160 mAh g$^{-1}$ at C/10 to 75 mAh g$^{-1}$ at 2C (Figure 3i), where both the specific capacity and current rate were referred to the mass of cathode. Notably, the higher working voltage of LiMn$_{0.5}$Fe$_{0.5}$PO$_4$ as compared to that of LiFePO$_4$ may have a detrimental on the coulombic efficiency of the battery.[53]

**Figure 3**

We remark that the presence of LiNO$_3$ in the formulation significantly affects the stability of the electrode/electrolyte interface, as demonstrated in two consecutive studies.[54,55] Indeed, the cycle life of Li|LiFePO$_4$ cells with PEGDME (MW 500 g mol$^{-1}$) dissolving LiTf was significantly enhanced by the addition of LiNO$_3$ to the electrolyte solution, leading to a stable performance over 60 cycle with a capacity referred to the cathode of 150 mAh g$^{-1}$ and a flat working voltage of 3.5 V, which were reflected as a theoretical energy density of 520 Wh kg$^{-1}$ (normalized to the mass of positive electrode). Furthermore, PFG-NMR measurements suggested suitable ionic conductivity, lithium transference number, ionic-association degree, and self-diffusion coefficient for energy-storage application. It is worth mentioning that the high thermal stability of this electrolyte may remarkably improve the safety of the lithium-metal battery.[54] In agreement with the above described results, a $G_3$–LiTf solution was upgraded by adding $LiNO_3$, which enhanced the electrode/electrolyte interface and widened the electrochemical stability window of the solution, thus enabling the application in a battery with the $LiFePO_4$ cathode. An electrochemical activation procedure of this cell leading to the formation of stable interfaces at the electrodes surface was optimized and thoroughly investigated. Figure 4 shows the related voltage profiles (a) and the corresponding SEM images of the $LiFePO_4$/electrolyte interphase upon cycling (b).[55] Unsymmetrical glymes having different end groups were also studied (Figure 4c). These glymes had ethyl and butyl end groups and were used to prepare liquid solvates with LiTFSI and LiFSI (glyme:salt ratios of 1:1, 2:1, and 3:1), which exhibited high ionic conductivity at room temperature (in the order of $10^{-3}$ S cm$^{-1}$), were stable up to 4.5 V, and ensured a steady galvanostatic performance with $LiFePO_4$, with a capacity of 145 mAh g$^{-1}$ (as referred to the cathode) and good rate capability up to 2C at room temperature. Furthermore, preliminary cycling tests with NMC suggested possible applicability in high-voltage batteries and DSC studies indicated a low crystallization temperature between −60 °C and −75 °C for the LiFSI-based electrolyte.[56] Therefore, both the salt and the glyme chain length may substantially affect the ion transport, the lithium/electrolyte interphase characteristics, and the electrochemical stability window, thereby determining the lithium cell response. Indeed, solutions of LiFSI, LiTFSI, or LiBETI in $G_n$ ($n$ = 2 and 3) had different properties depending on the selected formulation. Thus, decreasing the chain length increased the ionic conductivity from ca. $10^{-3}$ to ca. $10^{-2}$ S cm$^{-1}$, despite having detrimental effects on the lithium transference number. Galvanostatic lithium stripping-deposition and EIS measurements (Figure 4d–i) suggested that all the electrolytes may form an interphase at the metal anode suitable for few charge/discharge cycles, with low values of polarization and rather constant resistance values. However, the use of LiBETI led to poor lithium-passivation properties over long-term cycling, whilst widening the anodic stability window to 4.6 V $vs$ Li$^+$/Li. Among these various formulations, LiTFSI-based electrolytes were the most adequate formulations for application in Li|LiFePO$_4$ batteries that ensured a capacity between 134 and 144 mAh g$^{-1}$ in a galvanostatic tests over 100 cycles at a C/3 rate (1C was 170 mA g$^{-1}$; both capacity and C-rate were referred to the mass of LiFePO$_4$ in the cathode).[57] The electrochemical performance of the Li|LiFePO$_4$ battery was further enhanced as described above, by adding LiNO$_3$ to these solutions. A thorough investigation of such improved solutions showed evidence of fast ion transport, wide stability window, suitable lithium-metal passivation, and cathode/electrolyte interphase characteristics that depended on the G$_n$ chain length ($n$ = 2 and 3). An optimized Li|LiFePO$_4$ battery using a G$_3$-LiTFSI-LiNO$_3$ electrolyte delivered 154 mAh g$^{-1}$ at C/3 without any decay after 200 cycles, as well as a retention above 70% after 500 cycles at 1C and 5C with a coulombic efficiency approaching 100% (Figure 4j), which benefit from a stable, ionically-conductive electrode/electrolyte interphase (Figure 4k–n).[59] Relevantly, a solid composite-electrolyte based on PEGDME (MW 2000 g mol$^{-1}$), dissolving LiTFSI and LiNO$_3$, and incorporating nanometric silica (SiO$_2$) particles was prepared by solvent casting and investigated in Li|LiFePO$_4$ polymer batteries. This electrolyte had an ionic conductivity higher than 10$^{-4}$ S cm$^{-1}$ at temperatures above 40 °C after subsequent heating and cooling cycles, exhibited a lithium transference number ranging from 0.22 at 45 °C to 0.27 at 70 °C and a low interphase resistance at the interphase with lithium metal, and was stable up to ca. 4.4 V. When used in the Li|LiFePO$_4$ battery at 50 °C, this composite electrolyte enabled a coulombic efficiency of about 100%, a capacity retention approaching 99% after 300 cycles at a C/3 rate, and a maximum capacity of 150 mAh g$^{-1}$.[58] Furthermore, G$_2$ and G$_3$ dissolving LiTFSI and LiNO$_3$ in concentration approaching the solvents saturation limit were used in lithium cells employing LiFePO$_4$ cathode, which exhibited a promising cycling performance. Notably, an additional reduction step at low voltage cutoff (i.e., 1.2 V) during the first discharge allowed the formation of a suitable SEI, as mentioned above (see Figure 4a and b), thereby leading to a coulombic efficiency of ca. 100%, a capacity approaching 160 mAh g$^{-1}$, and low capacity fading over cycling.[60]

**Figure 4**

We can reasonably evaluate the glyme-based electrolytes (in particular using $G_3$ and $G_4$ and PEGDME) as possible solvents for electrolyte for lithium battery application, since they can have a stable electrode/electrolyte interphase, particularly when *ad-hoc* additives are used, low volatility, and sufficient dissociation degree. In particular, efficient operation is observed with typical electrodes used in battery such as those based on Li-intercalation ($LiCoO_2$ and graphite layered materials) and Li-insertion ($LiFePO_4$ olivine and $Li_2Ti_5O_{12}$ spinel). Possible formation of SILs due to salt concentration change is suggested to remarkably improve the oxidative stability of glyme electrolytes up to 5V vs. $Li/Li^+$, and the migration of G-S complexes in specific formulations is indicated to improve the cell performances, depending on the current value which is the limiting factor due to the depletion of the lithium salt in the electrode proximity. New electrodes such as $LiNi_{1/3}Mn_{1/3}Co_{1/3}O_2$ and $LiMn_{0.5}Fe_{0.5}PO_4$ can be also employed in lithium cells using glyme-based electrolyte with a stability and efficiency depending of the upper voltage cutoff. All the measurements indicate that the control of the salt concentration at the electrode/electrolyte interphase during cell operation, and the formation of a favorable SEI are the key factors for achieving the optimal operation of the lithium cells using insertion or intercalation electrodes. In particular, the addition of $LiNO_3$ to the electrolyte formulation as a sacrificial film-forming agent appears to significantly improve the SEI at the electrodes, in particular using glymes having the longer chains such as PEGDME and adopting suitable electrochemical activation steps. We remark again that the glyme electrolyte ionic transport, electrochemical stability and interphase characteristics are substantially governed both by salt nature and solvent chain length, and by the specific combination and experimental setup used for allowing the efficient cell operation. Therefore, targeted studies are suggested for investigating the various systems and further evaluate the actual battery applicability.

**3. Lithium salts in glymes: ionic liquids or concentrated solutions?**

Electrolytes formed by dissolving various lithium salts in glymes differing by the ether chain length were initially considered as simple solutions, mainly characterized by the solute concentration.[16–24]

Subsequently, the concept of liquid solvated salt-complexes with similar characteristics to those of ionic liquids, i.e., molten salts, was proposed.[133] In this regard, Raman spectroscopy may shed light on the various solvate structures occurring in the liquid phase at room temperature, whilst PFG-NMR may reveal information on the ion transport properties that can be crucial to understand the ion-solvent interactions. Accordingly, a literature work investigated solutions of LiFSI in $G_3$ or $G_4$, where the $G_n$:salt molar ratio was 1:1, which were identified as SILs comprising a cationic $[Li(G_n)]^+$ complex and the $FSI^-$ anion based on Raman data. For $G_n$:LiFSI ratios higher than 1, anionic $Li_xFSI_y$ complexes were formed in addition to the cationic one. Moreover, PFG-NMR revealed that the self-diffusion coefficients of $Li^+$ ($D_{Li}$) and glyme ($D_{glyme}$) were the same when $G_n$:LiFSI was 1:1, which indicated that $Li^+$ and glyme diffuse together as a cationic $[Li(G_n)]^+$ complex. The ratio of the self-diffusion coefficients of anion and cation, $D_{FSI}/D_{Li}$, was constant at ca. 1.1 – 1.3 when $G_n$:LiFSI was 1:1 and increased as the amount of LiFSI in the solution was raised, suggesting a change in the ion transport mechanism. Furthermore, the increase in LiFSI concentration enhanced the oxidative stability of the electrolyte and mitigated the Al corrosion.[61] The local structure of $Li^+$ ions in equimolar mixtures of glymes ($G_3$, $G_4$) was investigated in solutions of LiTFSI, LiBETI, LiTf, $LiBF_4$, $LiClO_4$, $LiNO_3$, and LiTFA. Raman spectra and *ab initio* molecular orbital calculations revealed a crown-ether like conformation of the glyme molecules to form a monomeric $[Li(G_n)]^+$ complex in the molten state. Raman spectroscopic data identified the fraction of the free glyme in the $[Li(G_n)]X$ solutions (were X is the anion; see Figure 5a), which was estimated to be a few percent in $[Li(G_n)]X$ with perfluorosulfonylimide type anions. Equimolar mixtures characterized by a low concentration of free glyme were regarded as SILs, while those containing a substantial amount of free glyme were classified as concentrated solutions. It is worth mentioning that the concentration of free glyme decreases as the concentration of salt increases, leading to a notable raise in the lithium-metal electrode potential. The significantly high electrode potential in SILs suggested the presence of stable $[Li(G_n)]^+$ complexes in the molten state.[62] In another work, the $[Li(G_4)_1][TFSI]$ SIL was mixed with various polymers, i.e., PEO, PMMA, and PBA, to prepare quasi-solid electrolytes for lithium batteries. The stability of the $[Li(G_4)]^+$ complex in the various solutions was assessed by comparing the ratio of the self-diffusion coefficient of glyme and $Li^+$ ion ($D_G/D_{Li}$). Thus, the $[Li(G_n)_1][TFSI]$ solution and the composites based on PMMA and PBA behaved as ILs since $D_G/D_{Li} = 1$ ($Li^+$ and $G_n$ diffuse together), whilst the PEO-based electrolyte was characterized by $D_G/D_{Li} > 1$, indicating the existence of free $G_4$ molecules (ligand exchange between $G_4$ and PEO). The highly stable $[Li(G_4)_1]^+$ complex (PMMA- and PBA-based solutions) led to high thermal stability, high $Li^+$ transference number, and wide electrochemical stability window. Among the investigated composites, the PBA-based solutions showed the lowest glass transition temperature, little affinity towards $Li^+$ ions, and favorable lithium transport properties.[63] SAXS, Raman spectroscopy, and computational modeling provided further insight on the structure of $G_4$–LiTFSI SILs. Indeed, a peak at Q 0.95 Å$^{-1}$ in the SAXS spectra indicated structural correlations of typical ILs. This peak grew in intensity as the concentration of salt increased, reaching a maximum at the equimolar ratio ($G_4$:LiTFSI 1:1; see Figure 5b) due to the effective solvation of each $Li^+$ ion by one $G_4$ molecule forming a $[Li(G_4)]^+$ complex. These data were confirmed by Raman spectroscopy analyses, which suggested the occurrence of SSIPs in the solution. However, it is worth considering that even at the equimolar concentration not all $Li^+$ ions were solvated and minor interaction between cation and anion was observed.[64] A further report suggested that changing the $G_3$:LiTFSI ratio may significantly affect the electrode/electrolyte interphase and, therefore, the cycle life of the cell. Indeed, SILs containing excess LiTFSI ensured notable mitigation of detrimental reactions between $LiCoO_2$ cathode and electrolyte.[65]

Glyme-based SILs typically exhibit a high viscosity, which may adversely impact the ion mobility, so various solvents additives able to enhance the conductivity of the solution have been studied. Among the diluents proposed so far, HFE demonstrated suitable properties for use in the battery. Hence, the liquid structure of HFE diluted $[Li(G_4)]_1TFSI$ was investigated combining Raman spectroscopy and DFT data, which suggested that the additive does not directly coordinate the $Li^+$ ion. SSIPs along with monodentate and bidentate CIPs were found in the neat and HFE diluted $[Li(G_4)]_1[TFSI]$, and the monodentate CIP decreased whilst the SSIP increased when diluting with HFE. HEXTS experiments were performed and supported by data of MD simulations, where the intermolecular force-field parameters, mainly partial atomic charges, were newly proposed for the HFE and glymes. A new peak ascribed to the correlation between the [Li(G$_4$)][TFSI] ion pairs was found at ca. 0.6 – 0.7 Å$^{-1}$ in the X-ray structure factors, and it was suggested that the terminal oxygen atoms of G$_4$ in the [Li(G$_4$)]$^+$ complex frequently repeat coordinating/uncoordinating, although almost all of the G$_4$ molecules coordinates the Li$^+$ ions.[66] The effects of dilution with various molecular solvents on the viscosity and the ionic conductivity of [Li(G$_n$)]$_1$[TFSI] molten complexes was further studied. Nonpolar solvents, such as toluene, DEC, and HFE, formed stable [Li(G$_3$)]$^+$ and [Li(G$_4$)]$^+$ solvates, whilst ligand exchange indicating competitive solvation took place between glyme and polar solvents, such as water and PC. On the other hand, ACN exhibited intermediate properties, as it participated in the lithium solvation to form mixed [Li(G$_3$)(ACN)]$^+$ and [Li(G$_4$)(ACN)]$^+$ complexes. Furthermore, [Li(G$_4$)][TFSI] was found to be more conductive than [Li(G$_3$)][TFSI] when diluted with nonpolar solvents due to a higher ionic dissociativity.[67]

Natural abundance $^{17}$O NMR spectroscopy provided information on the solvation behavior of LiTf and LiTFSI in glymes with different chain length, that is, G$_n$ with $n$ = 1, 2, 3, and 4 and PEGDME with MW of 250 and 500 g mol$^{-1}$. The chemical shifts of the glyme oxygen in the solutions and in the neat solvents was compared, revealing a more pronounced effect of the salt addition on the chemical shift of ether oxygens compared to that of terminal oxygens, which suggested a preferential coordination of Li$^+$ with the ether oxygen. The NMR data showed a mitigation of the chemical-shift changes as the glyme chain length increased (see Figure 5c), which was attributed to the increased number of ether oxygens coordinating each Li$^+$. Moreover, the chemical shift of anion oxygen suggested that the long chain length may decrease the ion association.[68] The solvation structure of lithium ions in the G$_3$–LiTFSI was determined performing MO and MD simulations based on DFT. According to these analyses, the Li$^+$ ions in the equimolar mixture were solvated mainly by crown-ether-like curled G$_3$ molecules and in direct contact with an TFSI$^-$ anion (see the snapshot in Figure 5d). The aggregate formed with Li$^+$ and TFSI$^-$ anions and/or G$_3$ chains was equally stable, thereby suggesting that a small fraction of cation may form aggregates.[69] Several characteristics of the anion, such as its flexibility and its ligand properties, influence the solvate structure, as demonstrated in a computational investigation of glyme-based solutions of LiTFSI and LiTDI. The related results revealed that TFSI$^-$ anions favor the formation of SILs, whilst TDI$^-$ anions preferably give rise to ionic aggregates. In particular, the latter anions ensured the presence of "free" cationic species even at extremely high salt concentration.[70] Important parameters may be extracted from PGSE-NMR measurements to provide a large dataset describing various electrolyte solutions. For example, a comparative analysis of several systems formed by dissolving LiPF$_6$, LiBF$_4$, LiTFSI, LiBETI, LiBOB, LiTf or Li$_2$DFB in PC, EC, GBL, DEC, G$_n$ (where $n$ = 2, 3, 4, and 5), or PEGDME of average MW of 400 and 1000 g mol$^{-1}$ was carried out by plotting the Li$^+$ and anion diffusion constants ($D_{Li}$ and $D_{Anion}$) versus the ionic conductivity ($\sigma$). The Nernst-Einstein (NE) relation was employed to calculate the degree of apparent ion dissociation ($\alpha$) from the $D_{Li}$, $D_{Anion}$ and $\sigma$ parameters. Furthermore, the apparent lithium transference number ($t_{Li}$) was determined the $D_{Li}$ and $D_{anion}$, and the number of charge-carrying ions ($N_{carrier}$) was estimated from $\alpha$ and the salt concentration. Additional relations were considered, such as $\sigma$ vs $\alpha$, $\sigma$ vs $N_{carrier}$, and $\sigma$ vs $t_{Li}$.[71] The Li$^+$ transport in glyme-based electrolytes was also investigated by comparing the characteristic features of various solutions, that is, 1.0 M LiTf, LiTFSI, or LiFSI in G$_4$ as well as 0.5, 2.0, or 2.7 M LiTFSI in G$_4$. Ionic conductivity ($\sigma$), viscosity ($\eta$), and density were measured, and the self-diffusion coefficients ($D$) of Li$^+$, anions, and solvent were determined via PGSE-NMR. This study suggested that the mobility ($\mu$) in the solution is controlled by the salt, as the Lewis basicity and hardness of anions affect the Li$^+$–anion and Li$^+$–G$_4$ interactions, as well as $\eta$ and $D$. The interaction energies ($\Delta E$) determined by DFT calculations based on the supermolecule method were found to be in the order LiTf > LiTFSI > LiFSI, which is consistent with the dissociation degree of these salts in solutions. The study demonstrated that increasing $\mu$ and number of carrier ions ($n$) are effective ways to enhance $\sigma$ for glyme-based electrolytes with low dielectric constant. In this regard, suitable properties may be achieved using DME.[72] VLF-EIS analyses of Li|Li symmetrical cells and MD simulations using atomistic,

polarizable force field provided additional insight into the Li$^+$ transport of equimolar SILs, enabling to calculate the three Onsager coefficients. Accordingly, a recent study of the G$_4$–LiTFSI mixture (1:1 ratio) showed that even though the ionic conductivity and the Li$^+$ transport number (i.e., the transport parameter extracted from PFG-NMR data) were acceptably high, the Li$^+$ transference number under anion-blocking conditions was extremely low. This observation was related to the strong complexation of Li$^+$ by glyme molecules, which leads to a long residence time of the solvent near the cation and causes significantly anti-correlation of cation and anion motion due to the constraint of momentum conservation (see Figure 5e). Moreover, it is worth mentioning that any transport number determined by PFG-NMR would substantially differ from the relevant electrochemical transference number in the presence of significant ion pairing. We also remark that strongly anti-correlated motion of cation and anion is typical of ionic liquids and, according to the structural model described above, glyme molecules complexed with Li$^+$ cannot ensure the momentum exchange between ions. Therefore, reducing the residence time between Li$^+$ and solvent molecules or diluting the solution (adding excess solvent) may increase the Li$^+$ transference number, as depicted in Figure 5e.[73] On the other hand, an unusual LiTFSP salt in a salt-in-glyme-based (salt-in-solvent) electrolyte solution exhibited predominant Li$^+$ conductivity with very high lithium transference numbers (70% from the polarization experiments), as well as an ionic conductivity that was three times higher than that of a solution of LiTf in G$_2$. This salt-in-solvent electrolyte was studied by PFG-NMR and EIS in symmetrical Li|Li cells, which suggested a suitable lithium-conduction behavior for applications in a battery, which was ascribed to the reduced mobility of large, solvated anions along with improved ionic dissociation.[74]

Recently, FPMD simulations identified the solvate structure in G$_4$–LiTFSI electrolyte solutions. If salt and solvent were in equimolar ratio, the simulations showed a positive correlation between the total coordination number of Li$^+$ ions and the phase stability. At the ground state of the equimolar G$_4$–LiTFSI mixture, curled G$_4$ molecules and TFSI$^-$ anions coordinated most of the Li$^+$ ions through 4 and 1 O atoms, respectively, which is the second most stable [Li(G$_4$)$_1$][TFSI] CIP in gas-phase cluster calculations. If the concentration of LiTFSI was low, Li$^+$ ions were not in direct contact with TFSI$^-$ anions and coordinated by two G$_4$ molecules. As expected, pairing between the Li(G$_4$)$^+$ and TFSI$^-$ was likely to occur at the equimolar ratio between salt and solvent, leading to properties typical of ionic liquids.[75] Cell temperature and salt concentration were shown to have effects on the ionic conductivity and the Li$^+$ transference number of liquid-solid, glyme-based electrolytes containing nanoporous alumina. Beneficial adsorption of the anion on the surface of the alumina particles had beneficial impacts on the cation transference number, mostly at lower salt concentration and elevated temperature. Even though the lithium transference number was high, unfavorable mossy lithium deposits was observed.[76] In this regard, the performance of the lithium anode was investigated also by comparing the effect of concentration, ether chain length, molecular structure, and electrolyte formulation of various solutions, namely, G$_n$–EC–LiFSI (where *n* = 1, 2, 3, and 4), DOL–EC–LiFSI and DEC–EC–LiPF$_6$. Conventional carbonate-based electrolytes using LiPF$_6$ gave rise to needle-like lithium dendrites, whereas high-concentration ether-based solutions favored knot-like and rounded lithium structures. Enhanced Li$^+$–solvent interactions and less free molecules were found in high-concentration EC-based electrolytes. These latter solutions showed less side reaction with the lithium-metal anode along with significant mitigation of the dendrite formation.[77] A recent study suggested a dynamic chelating effect, which was found to be related to solvent exchange and/or contact ion-pair formation/dissociation, having significant effects on the Li$^+$ transport. Data on the pair correlation functions for the Li–O(G$_n$) and Li–O(TFSI$^-$) pairs enabled to propose the PMF trend depicted in the schematic of Figure 5f, which suggested the predominance of CIP and Li$^+$–glyme complex formation for short-chain and long-chain solvents, respectively. According to this work, G$_4$ had intermediate characteristics, showing solvent exchange and CIP association/dissociation at a similar rate, which ensures high conductivity and low viscosity.[78] In agreement with these observations, a comparison of NMR diffusion and EIS data on glyme-based solutions of LiPF$_6$ revealed stronger ion pairing with decreasing glyme chain length. Furthermore, a decrease in the dielectric constant of the solvent with increasing temperature was suggested to further increase the ion association, thereby adversely affecting the ionic conductivity.[134]

**Figure 5**

As mentioned above, the lithium transference number in glyme-salt mixtures is influenced by ion-ion anticorrelations. This aspect was further demonstrated by eNMR measurements, which revealed the migration of the species forming $G_4$–LiTFSI or $G_4$–LiBF$_4$ electrolytes in an electric field. Relevant data on the electrophoretic mobility and the self-diffusion coefficients for the $^1$H, $^7$Li, and $^{19}$F nuclei provided insight on transference numbers, effective charges, and ionicities in various solutions characterized by a different salt concentration. In agreement with other reports, eNMR suggested that the $G_4$ molecule migrates along with the cation due to formation of a stable solvate complex. Furthermore, NMR showed an increase in effective charges as the solution approaches the equimolar ratio between glyme and salt and a notable difference in effective charge for lithium and anion, as well as between the $G_4$–LiTFSI and $G_4$–LiBF$_4$ systems. A schematic representation of the ion-ion and ion-solvent interactions in these solutions as a function of concentration and salt composition is reported in Figure 6a. Anticorrelations between solvate cation and the anionic complexes due to momentum conservation was identified as a crucial phenomenon possibly affecting the behavior of these mixtures in a different extent, depending on the employed electrolyte formulation. In this regard, high salt concentration along with the use of smaller anions that may form large asymmetric clusters may lead to strong anticorrelation.[79] Therefore, a clear understanding of the solvate structure may elucidate the Li$^+$ transport mechanisms in highly concentrated solutions, thereby providing crucial insight for electrolyte design. Concerning this point, a study of speciation in concentrated $G_n$–LiX and aqueous electrolyte systems identified the solvent activity and the activity coefficient in the gas phase at equilibrium with the solution as suitable parameters to classify the mixtures as SIL or superconcentrated solution. Thus, analyses performed with a Raman/IR spectral analysis technique, which may reveal the free-solvent concentration, found out that the solution can be regarded as SIL whether activity coefficient ($f$) is lower than 0.01. Accordingly, electrolyte solutions with the compositions of $(H_2O)_{1-x}$–LiTFSI$_x$, $(G_3)_{1-x}$–LiTFSI$_x$, and G$_3$–LiX (where X is $NO_3^-$, TFA$^-$, or TFSI$^-$) can be classified as shown in Figure 6b, by reporting $f$ as a function of $x$, that is, the salt molar fraction.[80] As mentioned, the Li$^+$ transference number estimated using the potentiostatic polarisation method of typical SILs, such as the G$_4$–LiTFSI equimolar mixture, is considerably lower than the corresponding transport number estimated via PFG-NMR measurements of the self-diffusion coefficient. This experimental evidence was interpreted by considering the dynamic ion correlations (i.e., cation–cation, anion–anion, and cation–anion cross-correlations) in the Onsager transport formalism, which describes the ionic conductivity $\sigma_{ion}$ as the given by equation (1):

$$\sigma_{ion} = \sigma_{++} + \sigma_{--} - 2\sigma_{+-} \tag{1}$$

where $\sigma_{++}$ and $\sigma_{--}$ are the transport coefficient of cation and anion, and the coefficient $\sigma_{+-}$ takes into account the cation–anion correlations. The former transport coefficients can be represented, in turn, by self-terms obtained from the self-diffusion coefficients of cations ($D_{Li}$) and anions ($D_{anion}$) applying the Nernst–Einstein equation (i.e., $\sigma_+^{self}$ and $\sigma_-^{self}$) and distinct terms ($\sigma_{++}^{distinct}$ and $\sigma_{--}^{distinct}$) calculated as per relations (2):

$$\sigma_{++}^{distinct} = \sigma_{++} - \sigma_+^{self} \quad \text{and} \quad \sigma_{--}^{distinct} = \sigma_{--} - \sigma_-^{self} \tag{2}$$

thereby leading to equation (3):

$$\sigma_{ion} = \sigma_+^{self} + \sigma_{++}^{distinct} + \sigma_-^{self} + \sigma_{--}^{distinct} - 2\sigma_{+-} \tag{3}$$

Notably, the sign of $\sigma_{++}^{distinct}$, $\sigma_{--}^{distinct}$, and $2\sigma_{+-}$ reflects the above-mentioned dynamic cross correlations. Therefore, the parameters calculated by normalizing the terms of equation (3) of to $\sigma_{ion}$ describe the contribution of these dynamic correlations to the ionic conductivity, as shown in Figure 6c for the [Li(G$_4$)][TFSI] system. In this equimolar mixture, cation–cation, anion–anion, and cation–anion migrations are anti-correlated, so that $\sigma_{++}^{distinct}$ and $\sigma_{--}^{distinct}$ negatively contribute to the ionic conductivity, whilst $2\sigma_{+-}$ positively contributes to the ionic conductivity. Hence, Figure 6c suggests that the momentum conservation of the $[Li(G_4)]^+$ complex and the $TFSI^-$ anion is achieved by momentum exchange of the ions, as typical of ILs. Notably, conventional concentrated solutions would instead exhibit significant momentum exchange between ions and solvent.[81] These interactions between the various species in the solution have major effects on the relation between the $Li^+$ transference number estimated via the potentiostatic polarization method ($t_{Li}^{PP}$) and the ionic conductivity (Figure 6d). Accordingly, high t$_{PPLi}$ and low ionic conductivity were observed in electrolytes containing anions with high Lewis basicity, which showed a strongly coupled, collective migration of $Li^+$ cations and anions forming clusters. An example of these solutions is the $[Li(G_3)][TFA]$ system, which is in fact characterized by all positive Onsager coefficients in equation (3), thereby leading to a $t_{Li}^{PP}$ as high as 0.90, in spite of an ionic conductivity below 0.1 mS cm$^{-1}$. On the other hand, the $\sigma_{++}^{distinct}/\sigma_{ion}$, $\sigma_{--}^{distinct}/\sigma_{ion}$, and $\sigma_{+-}/\sigma_{ion}$ coefficient were all negative for the $[Li(G_3)][TFSI]$ system, which is reflected as a relatively high conductivity and low $t_{Li}^{PP}$ (i.e., about 1.1 mS cm$^{-1}$ and 0.028, respectively). Furthermore, the five Onsager transport coefficients for $[Li(G_3)][TFA]$ were larger than those of $[Li(G3)][TFSA]$ by about two order of magnitudes due to the low ionic conductivity of the former electrolyte.[81,82]

**Figure 6**

The data reported above clarify in part the features of electrolytes formed by dissolving salts in glymes. Under specific concentrations the G-S systems as classified as liquid solvated salts (SILs), similarly to ionic liquids (ILs), in which both cations and anions contribute to the ionic conductivity. The free-glyme fraction is the parameter taken into account for rationalizing the G-S characteristics. Indeed, solutions with a low free-glyme fraction belong to the SILs, particularly using G$_3$ and G$_4$, whilst those with a high free-glyme fraction appertain to the concentrated solution class. The stability of $[Li(G_n)]^+$ complexes, strongly depending on the glyme chain length and the anion nature, affects the thermal stability and Li-transference number of the electrolyte. Furthermore, the formation of SILs can increase the electrolyte viscosity and hinder ion mobility and cell performances. The use of nonpolar additives in the glyme solutions can actually increase the ion mobility without altering the Li-solvent coordination and structure, which is instead changed by polar additives. In addition, the coordination of $Li^+$ ions to the ether-oxygen which has a key role to allow the formation of complexes is correlated with the salt concentration. For example, the formation of crown complexes and aggregates compatible with SILs is observed when salts having the $TSFI^-$ anion are used at low concentration. Interestingly, both mobility and viscosity of the glyme electrolyte are controlled by the Lewis basicity and hardness of the anions, while the number of ion carriers and the ionic conductivity increase by using glymes with relatively low dielectric constant. Moreover, the strong interactions between glyme chains and $Li^+$ occurring in concentrated solutions can increase the residence time of the solvent nearby the ions, thus decreasing the $Li^+$-transference number. Therefore, diluted solutions have typically a higher $Li^+$-transference number than concentrated ones, except for some particular case (e.g., the solvent-in-salt electrolytes in which very large anions are used to remarkably increase the $t_{Li+}$). The temperature affects the formation of SILs that, in turn, govern the growth of the lithium dendrites at the metal surface. Indeed, diluted solutions may favor the formation of noodle-like dendrites, instead knot-like and rounded lithium structures can be formed in concentrated solutions. Interestingly, predominant CIP formation and stronger ion pairing are observed in electrolytes using short-chain glymes compared to those using long-chain glymes, in which $Li^+$-glyme complexes are preferentially formed, whilst ion association degree is modified by changing the temperature value which directly affects the dielectric constant. Furthermore, ion-ion anticorrelation with migration of G-S clusters within the electric field is affected by the salt concentration and anion size, where high concentrations and small ions lead to large asymmetric clusters with strong anticorrelation. Most likely, the glyme solutions can be regarded as SILs whether the activity coefficient (f) is lower than 0.01, and the increase of the anion's Lewis basicity favors the lithium transference number, however depresses the ionic conductivity. Taking into account the incomplete scenario depicted by the several finding reported above, we can reasonably suggest further and more systematic studies aimed to fully rationalize the complex features of the glyme-based electrolyte since they are actually assuming an increasing importance in view of their possible application in a new generation of energy storage systems, as it will be illustrated in the subsequent paragraphs.

## 4. "Glyme electrolyte" in Li-S cell: a battery close to practical applications

Glymes dissolving lithium salts appeared as very promising candidates as lowly flammable electrolytes for application in efficient and high-energy Li-S batteries. These cells react via a multielectron conversion process delivering a theoretical specific capacity as high as 1675 mAh g$^{-1}$ (as referred to the mass of sulfur), which is reflected as a gravimetric energy density of ca. 3600 or 2600 Wh kg$^{-1}$ when referred to the mass of sulfur or Li$_2$S, respectively.[135] The electrochemical conversion of sulfur and lithium occurs at about 2.4 and 2.1 V $vs$ Li$^+$/Li according to relation (4):

$$16\text{Li} + \text{S}_8 \rightleftarrows 8\text{Li}_2\text{S} \qquad (4)$$

This reaction involves lithium polysulfide intermediates (Li$_2$S$_x$ with $2 \leq x \leq 8$), which dissolve in common electrolyte media for $x > 2$, giving rise to complex, potential-dependent equilibria between various species with different oxidation state.[136] Accordingly, Li-S cells undergo a gradual cathode loss in the electrolyte solution upon the electrochemical process, which may worsen the cycling performance and typically requires the optimization of a suitable cell design differing from that of conventional lithium-ion batteries. So far, a great deal of effort has been devoted to mitigating the detrimental effects of polysulfide dissolution, which has been regarded as one of the most challenging issues presently hindering practical applications of Li-S batteries. In this regard, the electrolyte formulation can radically affect the pathways of electrochemical reaction and, thus, the cell performance.[137]

A pioneering work reported the characteristic features of two lithium cells employing a pitch carbon-coated Li$_2$S cathode and a solid-state PEO$_{20}$LiTf–Li$_2$S–ZrO$_2$ electrolyte (LiTf/EO molar ratio of 1:20) or a G$_4$–LiTf liquid electrolyte (G$_4$/LiTf molar ratio of 1:4), respectively. The former battery delivered a stable capacity of 500 mAh g$_{\text{Li2S}}^{-1}$ when cycled at a C/3 rate and at 80 °C, while the latter battery exhibited a lower capacity, that is of about 300 mAh $g_{Li2S}^{-1}$, when tested at a C/6 rate and at 30 °C.[83] A more recent work reported a lithium-sulfur polymer battery using solid PEGDME (MW 2000 g mol$^{-1}$) operating at 50 °C with a capacity approaching 700 mAh $g_S^{-1}$ over 90 charge/discharge cycles. The polymer electrolyte has shown high thermal stability and stable interphase during the Li-S conversion process at 2.2 V vs. Li$^+$/Li.[104] Besides, [Li(G$_3$)$_4$][TFSI] and [Li(G$_3$)$_1$][TFSI] molten complexes were investigated as electrolyte solutions possibly able to mitigate the polysulfides dissolution. Indeed, a Li-S cell using the [Li(G$_3$)$_1$][TFSI] complex delivered a discharge capacity higher than 700 mAh $g_S^{-1}$ with a coulombic efficiencies above 98% over more than 400 cycles. Further evidence suggested that the addition of a nonflammable fluorinated solvent that preserve the solvate complex structure may improve the rate capability of the battery.[84] Following this trend, G$_n$–LiX equimolar mixtures (where $n$ = 3 or 4 and X is BETI$^-$, TFSI$^-$, Tf$^-$, BF$_4^-$, or NO$_3^-$) were investigated in Li-S cells and the dissolution of lithium polysulfides (Li$_2$S$_x$) was measured. According to this study, electrolytes exhibiting SIL characteristics such as [Li(G$_n$)][BETI] and [Li(G$_n$)][TFSI] may effectively ensure the suppression of the Li$_2$S$_x$ dissolution, as shown in Figure 7a. We remark that equimolar SILs would not contain "free" solvent molecules that can solvate the Li$_2$S$_x$ species, as widely discussed in the previous section. On the other hand, lithium polysulfides were highly soluble in concentrated solutions, such as G$_n$–LiTf and G$_n$–LiNO$_3$ (Figure 7a). Therefore, the batteries employing SILs ensured a stable response in galvanostatic cycling test, delivering a capacity between 600 and 700 mAh $g_S^{-1}$ with a coulombic efficiency above 98% over 100 cycles, whereas those using concentrated solutions displayed very poor performances. Moreover, the typical irreversible reduction of NO$_3^-$ anions at the positive electrode during discharge was observed, whilst BF$_4$ anions gave rise to detrimental side reactions with the polysulfide anions leading to undesired byproducts.[85] Alongside to these remarkable effects of the anions, the G$_n$ chain length was shown to have a significant influence on the mobility of the various species in the electrolyte solution, thereby remarkably impacting the Li-S cell behavior. Furthermore, the beneficial effects on the "polysulfide-shuttle" process of the addition of LiNO$_3$ along with LiTf to the electrolyte formulation were widely demonstrated. This adverse process consists in an apparent charging without actual energy storage due to the simultaneous oxidation of polysulfides at the cathode and their direct reduction at the anode upon migration in the solution across the cell separator. In this regard, Figure 7(b and c) shows that the ineffective charge of a Li-S cell employing a $G_2$–LiTf electrolyte (see panel b) is actually suppressed after addition of $LiNO_3$ to the solution (see panel c).[86] Another work proposed that dissociation of polysulfide dianions to radicals, and particularly the trisulfur radical ($S_3^{\cdot-}$), may play a crucial role in the electrochemical behavior of the Li-S cell. Accordingly, *operando* XANES measurements revealed these radicals in batteries using solutions of $LiClO_4$ and $LiNO_3$ in either $G_4$ or DMA, thereby suggesting that $S_3^{\cdot-}$ existed in minor concentration in glymes and in relatively high concentration in electron pair donor solvents. Notably, a high degree of dissociation of the anion precursor, $S_6^{2-}$, to the trisulfur radical in the latter solvents allowed the full utilization of both S and $Li_2S$. As shown in this study, catholyte-based batteries would benefit from low-volatility, electron pair donor solvents, and the combination of a highly adsorptive cathode with a highly dissociative solvent could be extremely effective in ensuring a superior performance. On the other hand, electron pair donor solvents would be detrimental for achieving polysulfide-entrapping in the cathode.[87] It is worth mentioning that the Li-S chemistry would affect also the lithium-passivation properties of the glyme based solution, as shown by a literature report which provided relevant XPS and electrochemical data on $G_4$-LiTf solutions charged by various polysulfide species, that is, $Li_2S_2$, $Li_2S_4$, $Li_2S_6$, and $Li_2S_8$. The results of this work demonstrated the presence of S-containing species in the SEI layer deposited over the lithium-metal anode and showed that the presence of polysulfides in the solution decreased the resistance of this layer. The addition of $Li_2S_x$ in the electrolyte stabilized the electrode/electrolyte interphase resistance and had a buffer effect which mitigated the cathode dissolution, leading to an improvement of the Li-S cell performance.[88]

Further approaches to reduce the polysulfide shuttling involved the design of electrolyte membranes with intrinsic microporosity, which would ensure size- and ion-selective transport. Polymeric membranes were optimized via MD simulations of the solvate structures of LiTFSI with lithium polysulfides ($Li_2S_x$, where $x$ = 8, 6, and 4) in glymes of different chain length. These simulations suggested that a pore size lower than 1.2 – 1.7 nm might bock the polysulfide crossover, thus possibly enabling a suitable electrochemical behavior of redox-flow, lithium-sulfur batteries, even in absence of $LiNO_3$ in the electrolyte formulation.[89] Alternative strategies recently explored to improve the cycle life and rate capability of Li-S cells included the addition of support solvents to SILs based on $G_n$–salt complexes. An example is represented by a novel fluorinated ether derivative, i.e., TFTFE, which was added to the $[Li(G_4)_1][TFSI]$ equimolar mixture, thereby enhancing the ionic transport across the electrolyte and, thus, the rate capability of the Li-S battery. Moreover, TFTFE was shown to decrease the solubility of lithium polysulfides in the electrolyte due to its low donor ability which limits the $Li^+$ solvation.[90] Another study comparing several equimolar mixtures of either $G_3$ or $G_4$ with various salts (LiBETI, $LiBF_4$, LiTf and $LiNO_3$) revealed that the dissolution of lithium polysulfides may be suppressed whether SILs are employed.[91] On the other hand, the $G_4$–LiTf solution with the lithium salt in a 1 mol $kg^{-1}$ concentration was considered the preferred electrolyte for performing a comparative evaluation of the electrochemical performance of several cathode composites prepared by mixing sulfur with carbon materials of various natures (namely, graphite, mesocarbon microbeads, and multi-walled carbon nanotubes). Besides, the electrolyte characteristics in terms of $^1H$, $^7Li$, and $^{19}F$ nuclei self-diffusion coefficients, ionic conductivity, and ionic association degree were investigated by combining NMR and impedance spectroscopy measurements. Notably, the best composite achieved in the Li-S cell using this electrolyte a capacity higher than 500 mAh $g^{-1}$ over 140 cycles, with no sign of dendrite formation or any shuttle reaction.[92] Glymes were also employed in combination with micropore-rich activated carbon incorporating sulfur,[93,96] as well as with S-Ketjenblack composites showing promising results.[94] Other cathode chemistries have been explored, such as a $TiS_2$–S composite reacting via intercalation of $Li^+$ ions into $TiS_2$ along with Li-S conversion.[95] Furthermore, a SIL consisting of a $[Li(G_4)][TFSI]$ equimolar mixture were investigated as a possible electrolyte for lithium-ion, silicon-sulfur batteries. In detail, pre-cycling in a FEC-containing solution before the test in the glyme-based electrolyte was shown to be an effective strategy to form a stable SEI incorporating fine LiF grains and organo-fluorine compounds over a Si-flake anode, which ensured a remarkable enhancement of the capacity retention.[97]

A recent work provided additional evidence of the effect of the lithium-salt concentration on the solubility of polysulfides in the electrolyte solution, which has been generally attributed to the presence of "free" glyme molecules favoring the solvation of $Li_2S_x$ species, as indeed mentioned above. In this regard, $[Li(G_n)]X$ complexes in equimolar mixtures (where LiX is a typical salt) can be locally destroyed to form "free" glyme chains upon the electrochemical process in Li-S cells. Therefore, non-equimolar mixtures (where the $G_n$:LiX molar ratio is higher than 1) may actually decrease the concentration of dissolved $Li_2S_8$, thereby improving the reversible capacity and the cycling stability, whilst increasing the coulombic efficiency. Photographic images (Figure 7d) and UV/vis spectra (Figure 7e) of various $Li_2S_8$-containing $[Li(G_3)_x][TFSI]$ SILs with $x$ of 0.8, 0.9, 1.0, 1.11, and 1.25, which were diluted with HFE, are consistent with a decrease in solubility of $Li_2S_8$ at increasing concentration of LiTFSI.[98] Although the mitigation of the dissolution of lithium polysulfides in the electrolyte has been regarded as a valuable strategy to enable high-performance Li-S batteries, several works have suggested that long cycle life and high coulombic efficiency may be achieved even in catholyte-based, semi-liquid configurations, provided that the lithium-metal anode is effectively protected against parasitic reactions with dissolved $Li_2S_x$. Catholyte solutions formed by dissolving either LiTFSI or LiTf along with $LiNO_3$ as anode-protection additive in $G_2$ were lately employed in Li-S cells exhibiting notable performances in terms of specific capacity, capacity retention, and coulombic efficiency. These solutions had suitable characteristics for applications, that is, satisfactory $Li^+$ transport ability, wide electrochemical stability window, and good Li-passivation properties. Semi-liquid cells with a sulfur loading ranging from 3 to 6 mg cm$^{-2}$ displayed a high rate capability, delivered a maximum capacity of ca. 1100 mAh g$^{-1}$ at a C/10 rate (Figure 7f), and ensured a stable capacity of about 800 mAh g$_S^{-1}$ at C/3 rate with coulombic efficiency exceeding 99% (Figure 7g).[99] Notably, the electrostatic attraction forces between $Li^+$ cation, salt anion, and $S_x^{2-}$ anion would play a significant role in determining the solvate structures in the electrolyte solution, thus significantly affecting the behavior of the cell. Test at low-temperature conditions shed light on this point, revealing that lithium polysulfides may form clusters that adversely impact the Li-S conversion kinetics. Therefore, a low testing temperature may lead to a poor performance, despite the low freezing point and high ionic conductivity of the glyme-based electrolyte. More abundant $Li^+$–$S_x^{2-}$ bonds and polysulfide clusters are formed whenever the cation as a higher affinity towards the $S_x^{2-}$ anion rather than the salt anion. On the other hand, it was suggested that a strongly bound lithium salt would disrupt polysulfide aggregates ensuring a substantially improved low-temperature performance. A graphical schematic of polysulfide clustering mechanism based on competing forces in the electrolyte solution is shown in Figure 7(h and i).[100] Another study of the cathode dissolution in the electrolyte during operation of the Li-S cell showed that lithium polysulfides are formed in both glyme-based and fluorinated-ether-based electrolytes, in which the polysulfides are highly soluble and lowly soluble, respectively. However, lithium polysulfides were trapped in the cathode pores when the latter solutions are employed, and the polysulfide concentration in the separator was below the limit of detection of HPLC analyses, thereby suggesting a decrease of interaction between polysulfides in solution in fluorinated-ether-based electrolytes. The different behavior of glyme-based and fluorinated-ether-based electrolytes affected the length and potential of the voltage plateaus of battery upon cycling.[101] The presence of $Li_2S_x$ in the electrolyte solution may have additional major effects on the cell response by influencing the composition of the SEI, as mentioned above and recently investigated.[102]

**Figure 7**

Li-S battery is presently considered the most energetic and promising energy storage system for future practical application. However, this alternative battery requires suitable electrolytes for allowing an adequate electrode/electrolyte interface and, at the same time, compatibility with lithium metal and low flammability. A further requested characteristic is the compatibility of the electrolytes with the polysulfides, which are unavoidably formed into the cell. Alongside with PEO-based electrolytes, those using glymes, in particular with the longer chain-length, appear the most promising candidates for achieving the above mentioned targets. Hence, solutions based on $G_2$, $G_3$, $G_4$ with the addition of $LiNO_3$ as the sacrificial film-forming agent to suppress polysulfide shuttle, as well as liquid and solid PEGDME can be successfully employed as the electrolyte media for Li-S cell. Interestingly, electrolytes with SIL characteristics above described appear to limit the polysulfide dissolution due to the absence of free solvent molecules, while polysulfide are soluble in more concentrated solutions. Furthermore, formation of radicals during the electrochemical process of the battery affecting the stability may be limited by using glymes with long chain. The presence of polysulfides in combination with $LiNO_3$ as the additive actually stabilizes the lithium metal interphase, and therefore solutions of polysulfides are proposed as the catholyte to operate in Li-S battery. It is worth mentioning that polymeric and porose membranes are also proposed with a certain success for mitigating the polysulfide shuttle effect even in absence of $LiNO_3$, however the use of the former is suggested alongside with the improvement of the cathode material. Lithium salt concentration and nature modify the structure of G-S and change the polysulfide clustering, thus remarkably affecting the performance of the cell. An interesting example is given by the non-equimolar G-S mixtures that decrease the solubility of the $Li_2S_8$ intermediate and enhance the battery stability, despite dissolved $Li_2S_8$ can be actually used as the liquid cathode in Li-S cell with $LiNO_3$.

## 5. "Glyme" electrolyte in Li-O$_2$ battery: the upcoming future

The Li-O$_2$ cell is an appealing next-generation battery system which may possibly ensure a notable breakthrough in energy density. This technology is based on the reaction (5) of oxygen with lithium in the electrochemical cell to produce lithium peroxide, occurring at 2.96 V *vs* Li$^+$/Li and delivering a formal specific capacity of 1168 mAh g$^{-1}$ as refereed to the mass of Li$_2$O$_2$.

$$2Li + O_2 \rightleftarrows Li_2O_2 \qquad (4)$$

However, the development of Li-O$_2$ batteries presents several challenges to overcome, which are mostly associated with parasitic reactions upon cell operation limiting the reversibility of the electrochemical process.[138,139] Furthermore, the carbonate-based electrolytes employed in conventional lithium-ion batteries decompose in contact with the electrochemical process intermediates, that is, superoxide and peroxide radicals,[140,141] so they are cannot be applied in Li-O$_2$ cells. On the other hand, excessively volatile ether solvents would pose serious limits for a system designed for potentially operating in open air.[142] In this regard, glymes are considered to be suitable solvents for Li-O$_2$ batteries,[143] since they have shown a high chemical stability as well as a boiling point and volatility that can be modulated by modifying the chain length.[19]

The applicability of G$_4$-LiTf in Li-O$_2$ cells was initially demonstrated by elucidating the reaction mechanism in a PEO-G$_4$-LiTf plasticized system. The results suggested that the highly solvating, base-resistant PEO plasticized by the low-molecular-weight, end-capped G$_4$ is a very good medium to study the electrochemical processes attributed to peroxide and oxide species in water-free Li-O$_2$ cells.[105] Subsequently, another work demonstrated that the use of a liquid G$_4$-LiTf electrolyte solution and an appropriate cell design can allow the Li-O$_2$ system to reversibly cycle at a current rate as high as 3 Ag$^{-1}$ delivering an outstanding capacity of 5000 mAh g$^{-1}$ (Figure 8a). As widely established in the Li-O$_2$ community, these values were referred to the mass of carbon in the electrode film, that is, Ag$_{carbon}^{-1}$ and mAh g$_{carbon}^{-1}$, respectively.[106] The electrolyte formulation significantly affects the electrochemical performance and the discharge products of Li-O$_2$ cells. Indeed, in a further study the use glymes as solvents led to a large amount of Li$_2$O$_2$ in the positive electrodes after discharge, while only a small amount of Li$_2$O$_2$ was detected after discharge in electrolytes based on nitrile, ionic liquid, phosphate, and sulfoxide. The employed solvent also influences the relative amount of Li$_2$CO$_3$ and LiF, which are formed as byproducts via oxidation and decomposition of the solvent and via the attack of superoxide radical anions to the binder and/or the F-containing imide salt, respectively. This work suggested the dibutyl diglyme as the most suitable solvent among those taken into account.[107] Therefore, glyme-based solution have been so far extensively investigated as electrolytes for Li-O$_2$ cell with promising results, and alternative chemistries have been also suggested. Accordingly, a pioneering report on the behavior of a lithium-ion-air battery where the metal anode is replaced by a lithiated silicon−carbon composite effectively employed a LiTf-$G_4$ electrolyte solution.[108] Besides the salt and solvent nature, the salt/solvent molar ratio plays a crucial role in the oxygen-conversion mechanism in the cell, as demonstrated by comparing a series of solutions of LiTFSI in $G_3$ and $G_4$ with various concentrations. According to this study, the $O_2^{·-}$ radical and solvated $Li^+$ would form an intermediate complex at the electrode/electrolyte interphase, that is, $Li^+(G_x)_n···O_2^{·-}$ (Figure 8b), leading in concentrated solutions (route a) to the formation of CIP solvates containing one $Li^+$ ion which is coordinated with four or five oxygen atoms of one glyme molecule. A decomposition pathway via reaction of $O_2^{·-}$ with the glyme was therefore proposed (route a in Figure 8b) to elucidate the performance degradations of Li-$O_2$ cells employing concentrated solutions. Dilution would gradually produce a mixture of (i) CIP and SSIP solvates, (ii) SSIP solvates (route b in Figure 8b), and eventually a mixture of SSIP solvates and free glyme molecules (route c in Figure 8b). Thus, intermediate concentrations would favor the accessibility of $O_2^{·-}$ to $Li^+$ and decrease the interactions between $O_2^{·-}$ and the glyme molecule that lead to detrimental decompositions (route b), whilst low concentrations would reasonably increase the frequency of collision between $O_2^{·-}$ and glyme, thereby enhancing the parasitic processes (route c). Therefore, the glyme/salt molar ratio of 1 to 5 was found to be critical for the cycling performance of the cell, and high stability over 20 cycles at 500 mA $g_{carbon}^{-1}$ were demonstrated using $(G_n)_5$−LiTFSI electrolytes with $n = 3$ and 4.[109] The electrolyte purity as another crucial aspect that should be taken into account, as shown in a qualitative and quantitative investigation of the reaction of $KO_2$ with glymes of various chain lengths via $^1$H NMR, FTIR, and UV-Vis spectroscopy, which demonstrated major effects on the cell performance.[110] Notably, glyme-based electrolytes would also ensure a wide temperature range of applicability, as demonstrated in a further report. As expected, the polarization and rate capability of a Li-$O_2$ cell using $G_4$–LiTf were shown to be influenced by the operating temperature (Figure 8c). Indeed, low temperatures would slow down the diffusion of $Li^+$ ions, whilst elevated temperatures would decrease the electrolyte viscosity and consequently increase the oxygen mobility. Thermal effects on the crystallinity of $Li_2O_2$ formed upon cell discharging were also observed.[111] It is worth considering that the reversibility of the Li-O$_2$ process upon full-capacity cycling is rather poor, and the outstanding long-term performances reported so far have been often obtained by limiting the capacity below 1000 mAh g$_{carbon}^{-1}$. On the other hand, it has been demonstrated that extended full-capacity cycling of Li-O$_2$ batteries using glymes would be ensured by selecting appropriate electrode materials (carbon source and catalyst) and cycling protocol. In particular, the formation of stable interfacial layer on the cathode surface during the initial cycling may stabilize the subsequent cycling stages. Whilst the initial cell operation was characterized by the predominant formation of Li$_2$O$_2$, the subsequent cycles led to the predominant formation of side products and to the eventual stabilization of the yield of Li$_2$O$_2$ at about 33 – 40%.[112] Besides, interesting results have been obtained by limiting the capacity of a Li-O$_2$ battery using an electrolyte with very low volatility to 1000 mAh g$_{carbon}{-1}$. The electrolyte was formed by dissolving LiTFSI in PEGDME (MW 500) and then mixing this solution with the Pyr$_{14}$TFSI ionic liquid.[113] Moreover, a new lithium-ether-derived chelate IL showed promising characteristics for applications in Li-O$_2$ batteries, namely, high stability against the lithium metal anode and against superoxide-initiated hydrogen abstraction when compared to DME. This electrolyte chemistry ensured a decrease in the amount of parasitic species formed during cycling, such as formate, as well as a ten-fold decrease in CO$_2$ evolution upon charge as compared to that observed in DME-based solutions.[114] Among the various possible lithium salts for aprotic solutions for Li-O$_2$ batteries, LiNO$_3$ dissolved in polyether solvents exhibited beneficial properties on both ORR and OER. The anion enhanced the former reaction by enabling the formation of submicrometric Li$_2$O$_2$ particles via a reaction pathway involving superoxide radicals and acted as redox mediator by producing at the negative electrode NO$_2^-$ ions which, in turn, formed NO$_2$ at 3.6 − 3.8 V *vs.* Li$^+$/Li. This latter species catalyze the Li$_2$O$_2$ oxidation forming NO$_2^-$, which is then oxidized again, thereby decreasing the OER overpotentials within the electrochemical stability window of glyme-based electrolytes.[115] 1 M solutions of LiTf, LiTFSI, and LiFSI in G$_4$ were further studied in Li-O$_2$ batteries to elucidate the interplay between ion transport and lithium stripping/plating, thereby revealing for the G$_4$–LiTFSI electrolyte mixture a high rate capability, a fast electrochemical reaction at the anode side, and a promising performance. Notably, the performance of the Li-$O_2$ cell depended on lithium oxide layers formed on the negative electrode. These results suggested that a suitable anode-surface oxidation may enhance the electrode/electrolyte interface characteristics.[116,120] The electrolyte formulation and the electrochemical conditions may affect both kinetics and thermodynamics of ORR, for instance by altering the solvation in high DN solvents. Along with solvent DN, the level of dissociation of the salt may have additional contributions in determining the characteristics of ORR and controlling the morphology of $Li_2O_2$ deposits, as schematically illustrated in Figure 8d.[117] Besides fundamental studies aiming to shed light on the electrochemical reactions occurring in the cell, further works focused on developing alternative battery designs, such as the metal-free Li-ion oxygen system which was mentioned above. For example, an effective approach to achieve excellent cycling stability and low cell polarization over 100 cycles involved the use of silicon particles at the negative electrode. In this regard, a stable SEI over the anode surface would mitigate oxygen crossover effects, which would improve the long-term cyclability of the battery.[118] Another lithium-ion oxygen cell using a lithiated hard carbon (HC) anode, a monolithic carbon cathode, and a $G_2$-based electrolyte was reported with promising results.[121] Additional evidence of the role played by the anion nature in the amount of $Li_2O_2$ precipitated on the separator in the Li-$O_2$ cell using glyme-based electrolytes suggested a relation between the stability of the discharge intermediate ($LiO_2$) in the electrolyte the anion. Accordingly, electrolyte solutions characterized by a high intermediate solubility would favor the precipitation of $Li_2O_2$ across the cell, thereby adversely affecting the reversibility of the oxygen conversion process. Figure 8e shows how the fraction of $Li_2O_2$ in the separator, as referred to the theoretical total amount of $Li_2O_2$, changes with the anion composition.[119]

As discussed in the previous sections, literature reports have described various beneficial properties of SILs when compared to conventional electrolytes for applications in lithium batteries employing insertion and sulfur-conversion cathodes. SILs may improve the Li-$O_2$ cell too, as demonstrated by a study of $G_3$–LiTFSI mixtures either where solvent and salts were in the equimolar ratio or containing excess glyme. The related results suggested that SILs would have a higher oxidative stability than that of conventional electrolytes, as well as a lower volatility, which would suit the open Li-$O_2$ cell configuration. High salt concentrations would mitigate the parasitic reactions that lead to $CO_2$ evolution, as shown in Figure 8f, although either type of solution did not ensure the full theoretical value of $O_2$ evolution, and side processes were detected upon charging. The discharge product morphology was also related to the solubility of the superoxide intermediate, which is in turn affected by the salt concentration.[122] Moreover, Raman spectroscopy analyses of glyme-based electrolytes suggested that the increase in LiTFSI concentration favors the formation of cationic and anionic complexes that stabilize the $G_4$ molecules against degradation. High-concentration electrolytes enabled an improvement as high as 300% in charge/discharge cycling tests with a limited capacity of 500 mAh g$^{-1}$, for solutions containing higher LiTFSI concentrations. A better cyclability at low $G_4$:LiTFSI molar ratio was associated with a decrease in growth rate of lithium carbonate species deriving from glyme degradation.[123] On the other hand, a $G_4$–LiTFSI solution with 1 mol kg$^{-1}$ concentration enabled to achieve hundreds of cycles without signs of decay to a Li-$O_2$ cell using a multiwalled carbon nanotube electrode. The reversibility of the electrochemical process in this cell was demonstrated by detecting the reversible formation and dissolution of $Li_2O_2$ during the electrochemical process.[124] Regarding this point, the $Li_2O_2$ oxidation pathway was associated with the solvent donicity according to a strong solvent-controlled mechanism. Thus, a solution route forming soluble $LiO_2$ intermediate after $Li_2O_2$ oxidation was suggested to occur in high-donicity solvents (Figure 8g, right-hand side pathway), whilst a solid-solution route forming solid $Li_{2-x}O_2$ intermediate was identified in low-donicity solvents (Figure 8g, left-hand side pathway). Notably, the former oxidation mechanism was causally related to an observed poor cycling stability of the relevant Li-$O_2$ cells.[125] A key aspect to consider when increasing the salt concentration in $G_n$–LiX solutions is the change in Li$^+$ transport properties and ionic conductivity, as indeed extensively discussed in section 3. We remark herein that viscosity and Li$^+$ transference number measurements on solutions using either $G_1$ or $G_2$ and either LiTf or LiTFSI have suggested a failure of the Walden's rule, in spite of a qualitative correlation with the association constant of the salts.[126]

The electrolyte composition may affect crucial parameters for applications in a Li-$O_2$ battery using a carbon-cloth GDL, such as the conductivity, viscosity, contact angle, and decomposition temperature. Among various formulations, the $G_4$–LiTFSI was often selected as the electrolyte of choice due to its suitable properties for use in the open metal-oxygen cell design. In this regard, a recent work demonstrated an enhanced performance as compared to those obtained with other formulations, that is, a longer cycle life at a discharge capacity limit of 2000 mAh g$^{-1}$ as referred to the mass of Pt catalyst in the GDL.[127] As mentioned above, LiNO$_3$-based solutions have been investigated as electrolytes for Li-$O_2$ batteries with promising results, which were attributed to the role of the NO$_3^-$ as redox mediator for the electrochemical process. Furthermore, LiNO$_3$ was shown to improve the lithium-electrolyte interphase by favoring the formation of a Li$_2$O layer which mitigates the lithium dendrite growth and the electrolyte decomposition. Dual solvent systems based on DMSO and $G_4$ were suggested as effective solutions characterized by an increased number per volume and mobility of Li$^+$ and NO$_3^-$. The former solvent has a relatively high dielectric constant and a low viscosity, which allowed a decrease in overpotential of the charge process along with an enhancement of the power density of the battery. Such a mixed formulation mitigated the low dissociation degree of LiNO$_3$ typically leading to low ionic conductivity.[128] As for the effects of the glyme chain length, SNIFTIRS and electrochemical data on Li-$O_2$ cells using gold electrodes lately suggested that $G_2$ may be more stable than $G_4$ between 3.6 and 3.9 V vs. Li$^+$/Li, although water impurities may reduce its stability.[129] On the other hand, a recent comparative study of concentrated solutions of LiTFSI and LiNO$_3$ salts in either $G_2$ or $G_3$ demonstrated that the former solvents would have volatility issues limiting its applicability in Li-$O_2$ cells with open design. Gradual solvent evaporation during cycling was observed along with a rapid cell failure. On the contrary, the use of the $G_3$-based electrolyte ensure stability over time as well as a specific capacity ranging from 500 to 1000 mAh g$_{carbon}^{-1}$. Such a concentrated $G_3$–LiTFSI–LiNO$_3$ was characterized by a high conductivity, a high Li$^+$ transference number, a wide electrochemical stability window, and favorable lithium-metal passivation properties.[130]

**Figure 8**

Despite still in a very early stage, the Li-$O_2$ battery is the most appealing energy storage system due to the highest theoretical potentialities among the various candidates. Parasitic reaction of the electrolyte solvents with the radicals formed during the Li-$O_2$ electrochemical process as well as the necessity of operating in an open environment (i.e., $O_2$ or dry air) represent the major drawbacks to overcome for enabling an efficient battery. Glyme electrolytes, which are modulable in terms of stability and volatility by changing the chain length, as well as the salt nature and concentration represent the most suitable electrolyte for allowing cell operation. Furthermore, the use of the lithium metal in a practical battery exposed to oxygen or air may be actually granted by long chain glymes such as highly viscous or solid PEGDME. In this regard, alternative lithium metal-free batteries are proposed in analogy with the Li-ion ones by using alloying electrodes at the negative side, however with a partial success. A solution based on $G_4$ represents the first reported example of an electrolyte in lithium oxygen cell with high capacity and stable trend. $G_3$ may be also adequate for Li-$O_2$ cell, instead $G_2$ and $G_1$ appear at the moment not suitable due to excessive volatility, and consequent evaporation during cell cycling in the open environment. Salt-to-solvent molar ratio is crucial for lithium-oxygen conversion mechanism: indeed, intermediate concentrations limit the glyme decomposition due to $O_2^{-\cdot}$ radical, instead low concentrations lead to detrimental decomposition. Decomposition reaction is in fact governed by the intermediate $Li^+(G_n)O_2^{-\cdot}$ Complex which can form coordinated $Li^+$ ions, CIP, SSIP solvates, and free glyme molecules. The latter species are principally formed at low concentrations and can react with $O_2^{-\cdot}$, thus leading to the electrolyte decomposition. In addition, solvent chemistry controls the reaction mechanism: solvents with high DN lead to the formation of soluble $LiO_2$ intermediate, instead solvents with low DN favor the formation of $Li_{2-x}O_2$ intermediate through a solid-solution mechanism. It is worth mentioning that the performances of the Li-$O_2$ cell are also affected by the temperature, the use of adequate GDLs, the cell setup and the experimental conditions, as well by the presence of $LiNO_3$ even though with less remarkable effects compared to Li-S battery. Furthermore, SILs formation may improve the Li-$O_2$ battery performance, while high salt concentration can mitigate the parasitic reactions due to the formation of cationic and anionic complexes, particularly using $G_4$.

**Remarks and conclusions**

Extensive works have been conducted over the past 25 years to assess the possible applicability of glymes in lithium batteries and shed light on the chemical-physical properties of their mixtures with lithium salts. Indeed, the various advantages of glyme-based electrolytes against common alkyl-carbonate-based solutions have been often highlighted, although some questions on their high-power performance when compared to their conventional, ester-based counterpart are still open. So far, glyme-based solutions have been thoroughly studied by using a wide portfolio of experimental methods, modelling approaches, and techniques, which have provided comprehensive description of their highly versatile characteristics, tunable by changing the glyme chain length along with the salt composition and concentration. Notably, glyme-based solutions typically exhibit a higher thermal stability as well as lower flammability and volatility than common lithium-battery electrolytes, particularly at middle and moderately high chain length, thereby possibly enabling the use of the high-energy lithium-metal anode. Thanks to such tailored properties, glymes can be effectively employed in lithium batteries with insertion cathodes working below 4 V, such as $LiFePO_4$, as well as with 4 V-class cathodes such as $LiCoO_2$, NMC and $LiMn_xFe_{1-x}PO_4$ when used in concentrated formulations as salt-in-solvent mixtures or SILs. Furthermore, glymes are especially suited for high-energy Li-S batteries and considered to be the solvents of choice for $Li-O_2$ batteries, due to a high chemical stability in the cell along with a suitable volatility for the open design. Among the various lithium salts for glyme-based electrolytes that have been investigated, LiTFSI, LiTf, LiFSI, and LiBETI have shown favorable characteristics in any of the above-mentioned cell designs based on insertion or conversion chemistries. Furthermore, the crucial role of $LiNO_3$ as film-forming additive to achieve reversible lithium-metal plating and anode protection in conventional Li-metal and Li-S batteries has been clearly demonstrated, and its additional redox-mediator properties in Li-$O_2$ cells has been suggested.

Interestingly, the electrolyte properties may vary within those of an SIL to those of a conventional electrolyte, depending on the concentration and nature of the salt, along with the glyme chain length. As extensively discussed in the previous sections, an SIL has been described as a [Li($G_n$)]X system, where Li$^+$ is mainly solvated by crown-ether-like curled $G_n$ molecules forming a complex in contact with the TFSI$^-$ anion, and negligible "free" solvent is detected in the mixture. In this regard, the stability of the [Li($G_n$)]$^+$ complex may be revealed by comparing the ratio between the self-diffusion coefficients of glyme and Li$^+$ ion ($D_G/D_{Li}$). Therefore, when designing a high-concentration glyme-based electrolyte, the peculiar features of the anion should be considered along with the solvent molecular weight. As an example, we remark that TFSI$^-$ would favor the formation of SILs, whilst TDI$^-$ would promote ionic aggregation. Herein, we point out that the Li$^+$ transference number of SILs under anion-blocking conditions is typically low because of the Li$^+$–$G_n$ complexation, which causes anti-correlation of cation and anion motion due to momentum conservation. On the contrary, conventional concentrated solutions exhibit momentum exchange between ions and solvent molecules. Strong Lewis base anions ensure collective migration of ion pairs, which leads to high a Li$^+$ transference number and a low ionic conductivity. Instead, anions with low Lewis basicity typically lead to a high conductivity and a low Li$^+$ transference number. The increase in viscosity might be another drawback of highly concentrated $G_n$–LiX mixtures, which may require the use of diluents to increase the ionic conductivity. Lithium dendrite formation and anode passivation properties are other key aspects to consider when optimizing an electrolyte formulation enhanced by additives.

Several works have suggested that SILs may improve the long-term cycling ability of Li-S cells by preventing the shuttle effect. According to this approach, the polysulfide dissolution would be significantly decreased by limiting the amount of "free" solvent molecules able to participate in the solvation process, thereby leading to a *quasi-solid* sulfur reaction mechanism in the cell. On the other hand, glyme-based catholytes have shown a promising performance in a wide range of current rates, benefiting from a highly reversible polysulfide conversion on the positive electrode and from suitable Li$^+$ transport properties. Concerning this point, the addition of LiNO$_3$ to the glyme-based electrolyte can actually mitigate the detrimental effects of polysulfide dissolution by enabling efficient protection of the lithium-metal anode against parasitic reactions. As for the Li-O$_2$ system, the electrolyte formulation controls the Li-O$_2$ conversion mechanism and the morphology of the Li$_2$O$_2$ discharge product, thus influencing the cell performance. Indeed, solvent DN and dissociation level of the salt may drive the pathway of ORR. Moreover, literature works have suggested favorable properties of SILs compared to conventional electrolyte solutions.

Figure 9 summarizes the main concepts discussed herein as well as the main conclusions drawn by reviewing the most relevant studies of glyme-based electrolytes for lithium batteries. The figure shows that the glyme-based electrolytes can allow the change from the intercalation chemistry to the more energetic and alternative conversion chemistry (top-side scheme) and the use of the lithium metal in battery (bottom-side scheme). In particular, PEGDME in the liquid and the solid state is nontoxic, nonflammable, safe and is can actually allow scalable Li-S and Li-O$_2$ battery for future applications. Fig. 9 also reveals that the increase of the salt concentration in the glyme-based electrolytes unlikely leads to the decrease of the lithium transference number, however it increases at the same time the safety and the stability of the battery (right-hand arrows). In this scenario the electrolyte viscosity may potentially represent advantages in terms of scalability, and disadvantages in terms of conductivity and cell performances. The fundamental investigation and applied research have indeed enabled a deep understanding of the G$_n$–LiX structure along with substantial technological improvements of the lithium cell. A critical analysis of the extensive experimental evidence collected during the last two decades demonstrates the suitability of glymes as electrolyte solvents for lithium-metal batteries, and suggests viable strategies to achieve enhanced, next-generation energy-storage systems.

**Figure 9**

**Acknowledgments**

This project/work has received funding from the European Union's Horizon 2020 research and innovation programme Graphene Flagship under grant agreement No 881603. The authors also thank grant "Fondo di Ateneo per la Ricerca Locale (FAR) 2021", University of Ferrara, and the Institute of Global Innovation Research (GIR) in Tokyo University of Agriculture and Technology.

**Tables captions**

**Table 1.** Acronyms of the various compound and chemical species cited throughout the text.

**Table 2.** Physical-chemical properties of carbonate and glyme solvents employed for battery application. The SDS data are provided by Sigma-Aldrich.

**Table 3.** Acronyms of the various techniques and expressions cited throughout the text.

**Figure captions**

**Figure 1. (a)** Schematic of the main research topics involving the fundamental and technological investigations of glyme-based electrolytes for lithium batteries carried out over the past 25 years. **(b-c)** Comparison between the present Li-ion battery and the two emerging energy storage systems (i.e., Li-S and Li-O$_2$ cells) which can be allowed by the use of glyme-based electrolyte for possible application in the electric vehicles in terms of **(b)** driving range (km by a single charge) and **(c)** economic impact (USD per kWh of battery pack).

**Figure 2. (a)** Typical steady-state voltammograms obtained from polycrystalline gold electrodes in LiTf 1 M solutions of G1 (DME), ethyl glyme (EG) and G$_2$ (DG). Scan rate: 20 mv s$^{-1}$. Li underpotential deposition and stripping peaks and the non-Faradaic region are marked. Reproduced with permission.[16] Copyright © 1996 Elsevier Science Ltd. **(b)** Specific conductivity ($\kappa$) at 25 °C of LiTFSI in glymes, that is, DME (EGDME; $\kappa$ at 35 °C is also shown), G$_2$ (2EGDME), G$_3$ (3EGDME), G$_4$ (4EGDME). Reproduced with permission.[17] Copyright © 1998 Plenum Publishing Corporation. **(c)** Relationship between electrolyte conductivity ($\kappa$), charge–discharge cycling current (I$_{ps}$) for maximum value of cycling efficiency of lithium (Eff$_{max}$) and $n$ for 1 M LiPF$_6$–EM$_{80}$/$n$G$_{20}$, where EM$_{100-x}$/$n$G$_x$ represents the mixed solvent of EC/MEC (3:7) and $n$-glyme (mixing volume ratio =100 − $x$:$x$); plating charge (Q$_p$) of 0.025 mAh. Reproduced with permission.[19] Copyright © 2003 Elsevier Ltd. All rights reserved. **(d)** Schematic illustrations of various glyme-LiX solvate structures with G$_1$, G$_2$, G$_3$, and G$_4$. Reproduced with permission.[21] Copyright © 2006 American Chemical Society.

**Figure 3. (a)** Temperature dependence of ionic conductivity of [Li(G$_3$)][CTFSI] and [Li(G$_4$)][CTFSI] glyme–Li salt complexes. Reproduced with permission.[43] Copyright © 2009 Elsevier B.V. All rights

reserved. **(b, c)** Linear sweep voltammograms of [Li(glyme)$_x$][TFSI] complexes ($x$ = 1, 4, 8, and 20) at a scan rate of 1 mV s$^{-1}$ at 30 °C, where the glyme is **(b)** G$_3$ and **(c)** G$_4$. Each inset depicts a magnification of current density. Adapted with permission (https://pubs.acs.org/doi/10.1021/ja203983r).[46] Copyright © 2011 American Chemical Society. Further permission related to the material excerpted should be directed to the American Chemical Society. **(d, g)** Electrochemical behavior of a Li|G$_3$–LiTFSI|LiFePO$_4$ cell at the 1st, 100th, 200th, 300th, 400th, and 500th cycles in terms of **(d)** voltage profiles and **(g)** trend of charge and discharge capacities per positive electrode as a function of the cycle number; temperature: 30 °C; voltage range: 2.5 – 4.0 V; current rate: C/8. **(e, h)** Electrochemical behavior of a Li|G$_3$–LiTFSI|LiNi$_{1/3}$Mn$_{1/3}$Co$_{1/3}$O$_2$ cell at the 1st, 10th, 50th, 100th, 200th, 300th, and 400th cycles in terms of **(e)** voltage profiles (voltage range: 2.7 – 4.2 V) and **(h)** trend of discharge capacity per positive electrode as a function of the cycle number for various upper charge cutoff voltages (4.2 V, 4.4 V, 4.6 V): temperature: 30 °C; current: C/8. Adapted with permission.[48] Copyright © 2013 The Authors. Published by Elsevier B.V. CC BY-NC-ND license. **(f, i)** Electrochemical behavior of the Li|LiTf–LiNO$_3$–G$_4$|LiMn$_{0.5}$Fe$_{0.5}$PO$_4$ cell in terms of **(f)** voltage profiles and **(i)** trend of charge and discharge capacities per positive electrode and coulombic efficiency as a function of the cycle number; temperature: 25 °C; voltage range: 2.0 – 4.3 V; current rate: C/5 rate (1C = 170 mAh g$^{-1}$ as referred to the cathode mass); test performed after an electrochemical activation of the cell. Reproduced with permission.[53] Copyright © 2016 Elsevier B.V. All rights reserved.

**Figure 4. (a, b)** Ex situ SEM investigation of LiFePO$_4$ electrodes after cycling in a Li|G$_3$−LiTf−LiNO$_3$|LiFePO$_4$ cells at the OCV condition (just after cell assembly and stabilization) and after the 1st, 10th, and 50th cycles; **(a)** voltage profiles and **(b)** SEM images. First discharge performed by decreasing the voltage below 2 V at C/5 rate and limiting the time to 5.15 h; subsequent cycles within the 2 − 4 V voltage range at C/5 rate (1C = 170 mA h g$^{-1}$ as referred to the cathode mass). Reproduced with permission.[55] Copyright © 2017 American Chemical Society. **(c)** Schematic

exemplifying the promising properties of solvates based on glymes with unsymmetrical (ethyl and butyl) end groups as electrolytes for lithium batteries characterized by high conductivity. Reproduced with permission.[56] Copyright © 2017 American Chemical Society. **(d–i)** Voltage profiles of lithium stripping/deposition galvanostatic cycling tests of symmetrical Li|Li cells and corresponding Nyquist plots of the EIS before the test (P1), after 5 cycles (P2), upon 24 h of rest after cycling (P3), and after additional 5 cycles (P4) for **(d)** $G_2$–LiTFSI, **(e)** $G_3$–LiTFSI, **(f)** $G_2$–LiFSI, **(g)** $G_3$–LiFSI, **(h)** $G_2$–LiBETI, and **(i)** $G_3$–LiBETI. Cycling test at a constant current density of 0.1 mA cm$^{-2}$. Step time: 1 h; EIS carried out by applying an AC signal with amplitude of 10 mV in frequency range from 500 kHz to 100 mHz; temperature: 25 °C. Reproduced with permission.[57] Copyright © 2019 Elsevier Ltd. All rights reserved. **(j)** Galvanostatic cycling trend in terms of discharge capacity and coulombic efficiency over 500 cycles of a Li|LiFePO$_4$ cell using the LiTFSI–LiNO$_3$–G$_3$ electrolyte at 1C, 2C and 5C rates (1C = 170 mAg$^{-1}$ as referred to the cathode mass). **(k–m)** Nyquist plots of EIS measurements performed on the same cell during the cycling tests at **(k)** 1C, **(l)** 2C, and **(m)** 5C rates; impedance spectra were recorded at the OCV condition, after the 1$^{st}$, 200$^{th}$, and 500$^{th}$ cycles. **(n)** Electrode/electrolyte interphase resistance for the same cell as extracted from the EIS data of panels **k–m**. Reproduced with permission.[59] Copyright © 2020 Wiley-VCH Verlag GmbH & Co. KGaA, Weinheim.

**Figure 5. (a)** Estimated percentages of free glyme ($c_f/c_G$, where $c_f$ and $c_G$ are the concentration of free glyme and the total concentration of glyme in the mixture, respectively) in equimolar molten mixtures [Li(glyme)$_1$]X at 30 °C (X is TFSI, Tf, NO$_3$, TFA, BETI, TFSA, ClO$_4$, and BF$_4$). Reproduced with permission.[62] This journal is © the Owner Societies 2015. **(b)** SAXS diffraction patterns of different mole fractions of LiTFSI dissolved in G$_4$. The dotted vertical lines represent guidelines to the eye. Reproduced with permission.[64] Copyright © 2015 Elsevier B.V. All rights reserved. **(c)** Chemical shift difference of $^{17}$O resonance of glyme between Li salt solution and pure solvent ($\Delta\delta = \delta_{electrolyte} - \delta_{solvent}$) as determined by NMR experiments. Adapted with permission.[68]

Copyright © 2016 Elsevier Ltd. All rights reserved. **(d)** Snapshot of the structure for the LiTFSI–$G_3$ electrolyte solution; CIP in which a Li ion is solvated by one $G_3$ molecule and one TFSI anion; the inset shows the representative solvation structures of the Li ion. Adapted with permission.[69] Copyright © 2016 American Chemical Society. **(e)** Schematic illustration of SIL with low $Li^+$ transference number due to momentum conservation constrain and two possible strategies to facilitate the momentum exchange in electrolyte and to increase $Li^+$ transference. Reproduced with permission.[73] This journal is © the Owner Societies 2018. **(f)** Schematic illustration of the potentials of mean force (PMF = − $RT$ ln $g(r)$, where $g(r)$ is the pair correlation functions g(r) for the Li–O atom pair) in glyme–$G_n$ solutions, suggesting the predominance of the CIP formation for short glyme chains ($n < 4$) and preference in the $Li^+$–solvate complexes for long glyme chains (n > 4). Reproduced with permission.[78] Copyright © 2019 American Chemical Society.

**Figure 6. (a)** Schematic representation of speciation: suggested dominant ion clusters in the low and high salt concentration regimes for $[Li(G_4)_x][TFSI]$ (top) and $[Li(G_4)x][BF_4]$ (bottom). Anionic species are depicted in blue, cationic species in red, and neutral species in gray. Reproduced with permission.[79] Copyright © 2020 American Chemical Society. **(b)** Plots of activity coefficients *vs* Li salt mole fraction. Blue and red denote $[LiTFSI]_x(H_2O)_{1-x}$ and $[LiTFSI]_x(G_3)_{1-x}$, respectively, and green denotes $[LiX][G_3]$ (X = $NO_3^-$, $TFA^-$, or $TFSA^-$); f < 0.01 is herein used as a thermodynamic criterion for categorizing the mixture as an SILs instead of a concentrated electrolyte solution. Reproduced with permission.[80] © 2020 American Chemical Society. **(c)** Five transport coefficients according to the Onsager formalism [$\sigma_+$(self), $\sigma_-$(self), $\sigma_{++}$(distinct), $\sigma_{--}$ (distinct), and $\sigma_{+-}$: left to right] normalized by the ionic conductivity ($\sigma_{ion}$) of the molten Li salt solvates for the $[Li(G_4)][TFSI]$ complex. Adapted with permission.[81] This journal is © the Owner Societies 2020. **(d)** Plots of the $Li^+$ transference number based on potentiostatic polarization measurements ($t_{Li}^{PP}$) *vs* ionic conductivity for $[Li(G_n)]X$ complexes (n = 3 and 4; X = TFA, $NO_3$, Tf, $BF_4$, NFSA, $ClO_4$, BETI, TFSI, and FSI). Reproduced with permission.[82] This journal is © the Owner Societies 2021.

**Figure 7. (a)** Solubility limits of $S_8$, $Li_2S_8$, $Li_2S_4$, $Li_2S_2$, and $LiS_2$, where $Li_2S_m$ ($m$ = 8, 4, 2) is the nominal formula of mixture prepared using the stoichiometric amounts of $S_8$ and $Li_2S$; the nominal formula assumes a complete reaction between $S_8$ and $Li_2S$ without occurrence of disproportion reactions; [Li(glyme)]X represented by empty and full symbols are categorized as SILs and concentrated solutions, respectively. Reproduced with permission.[85] Copyright © 2013 American Chemical Society. **(b, c)** Voltage profiles of Li-S cells using **(b)** $G_2$–LiTf (LiTf in a concentration of 1 mol $kg_{solvent}^{-1}$) and **(c)** $G_2$–LiTf–LiNO$_3$ (LiTf and LiNO$_3$ in concentrations of 1 mol $kg_{solvent}^{-1}$ and 0.4 mol $kg_{solvent}^{-1}$, respectively); temperature: 25 °C; voltage range: 1.2 − 3.2 V; current rate: C/20 (83.75 mA $g^{-1}$ as referred to the cathode mass). Adapted with permission.[86] Copyright © 2015 American Chemical Society. **(d)** Photograph and **(e)** UV–vis spectra of [Li(G$_3$)$_x$][TFSI] solutions of Li$_2$S$_8$ (x = 0.8, 0.9, 1.0, 1.11, and 1.25) diluted four times with HFE at room temperature. Reproduced with permission.[98] Copyright © 2019 WILEY-VCH Verlag GmbH & Co. KGaA, Weinheim. **(f, g)** Electrochemical behavior of a catholyte-based, Li|G$_2$–Li$_2$S$_8$–LiNO$_3$–LiTf|C cell in terms of **(f)** voltage profiles at C/10, C/8, C/5, C/3, C/2, 1C and 2C rates (1C = 1675 mA $g_S^{-1}$) and **(g)** trend of charge and discharge capacities per positive electrode and coulombic efficiency at a C/3 rate (1C = 1675 mA $g_S^{-1}$) as a function of the cycle number (both LiTf and LiNO$_3$ in concentrations of 1 mol $kg_{solvent}^{-1}$); voltage range: 1.8 – 2.8 V form C/10 to C/3 and 1.7– 2.8 V from C/2 to 2C. Adapted with permission.[99] Copyright © 2018 Elsevier B.V. All rights reserved. **(h, i)** Illustration of competing interactions between lithium species in electrolyte solutions for Li-S batteries; **(h)** strong Li$^+$–S$_x^{2-}$ bond networks disrupted from competing electrostatic interactions between lithium ions and lithium salt anions; **(i)** lithium polysulfides that naturally form at low temperatures, disrupted from the influence of competing lithium salt. Reproduced with permission.[100] Copyright © 2020 American Chemical Society.

**Figure 8. (a)** Voltage profiles of a Li|G$_4$–LiTf|O$_2$ battery under a 5000 mAh $g^{-1}$ specific-capacity limit as referred to the mass of carbon cast on the cathode GDL; current rate: 500 mA $g^{-1}$ (current

rate referred to the mass of carbon cast on the cathode GDL). Adapted with permission.[144] Copyright © 2012 Macmillan Publishers Limited. All rights reserved. **(b)** Mechanism of $O_2$ reduction reaction in a Li-$O_2$ cell at the interface between carbon black and glyme–LiTFSI electrolyte solution during discharge as proposed elsewhere. Reproduced with permission.[109] Copyright © 2013 WILEY-VCH Verlag GmbH & Co. KGaA, Weinheim. **(c)** Cycling behavior of TEGDME-based electrolyte lithium−oxygen cells at 25 °C (a), 50 °C (b), 70 °C (c), 0 °C (d), −10 °C (e). Voltage *vs* temperature trend for a Li-$O_2$ cell using a $G_4$–LiTf electrolyte solution. Insets: ex situ TEM images of the cathode of the same cell after galvanostatic discharge at 25 °C (left-hand side) and at 70 °C (right-hand side); capacity limited to 10000 mAh $g^{-1}$ as referred to the mass of carbon cast on the cathode GDL; current rate: 500 mA $g^{-1}$ (current rate referred to the mass of carbon cast on the cathode GDL). Reproduced with permission.[111] **(d)** Schematic representation of the effect of salt anion (i.e., $TFSI^-$, $Tf^-$, $Br^-$, and $NO_3^-$) in electrolyte solutions based on $G_2$ on the kinetics and thermodynamic of the ORR, as well as on the morphology of electrodeposited $Li_2O_2$, in Li-$O_2$ batteries. Reproduced with permission.[117] Copyright © 2016 American Chemical Society. **(e)** Results of quantitative determination of $Li_2O_2$ on the separator in a Li-$O_2$ cells using LiX/$G_3$ (1:4) electrolytes, where $X^- = BETI^-$, $TFSI^-$, $Tf^-$, $NO_3^-$, and $TFA^-$. The values on the left y-axis are the percentage of the amount detected in the separator with respect to the amount of total expected (theoretical) $Li_2O_2$ present in the cell after discharge to 1 mAh $cm^{-2}$, whilst the corresponding absolute values (μmol) are shown on the right y-axis; error bars denote the standard error. Adapted with permission.[119] Copyright © 2017 The Chemical Society of Japan. **(f)** Integrated $CO_2$ evolution during charge of Li-$O_2$ cells using either $Li(G_3)_1$][TFSI] (blue curve) or [Li$(G_3)_4$][TFSI] (red curve) electrolytes, which was calculated from electrochemical MS data, after galvanostatic discharge to 1 mAh. Adapted with permission (https://pubs.acs.org/doi/pdf/10.1021/acsami.6b14449).[122] Copyright © 2017 American Chemical Society. Further permission related to the material excerpted should be directed to the American Chemical Society. **(g)** Solvent-controlled $Li_2O_2$ decomposition mechanism in Li-$O_2$ batteries as proposed elsewhere; ''H'' denotes high-donicity solvent and ''L'' denotes low-donicity solvent.

Li$_2$O$_2$* denotes the Li$_2$O$_2$ generated by LiO$_{2(sol)}$ disproportionation, where LiO$_{2(sol)}$ is a soluble discharge intermediate product of the ORR. Reproduced with permission.[125] Copyright © 2018 Elsevier Inc. **(h)** Voltage profiles of a Li-O$_2$ cell using G$_3$–LiTFSI–LiNO$_3$ (both LiTFSI and LiNO$_3$ in concentrations of 2 mol kg$_{solvent}^{-1}$); specific capacity limited to 500 mAh g$^{-1}$ as referred to the mass of carbon cast on the cathode GDL; current rate: 100 mA g$^{-1}$ (current rate referred to the mass of carbon cast on the cathode GDL); carbon loading over the cathode GDL: 0.65 mg cm$^{-2}$ (carbon mass: 1.3 mg, geometric electrode area 2 cm$^2$) voltage range: 1.5 − 4.6 V; cell aged for 7 days before cycling. Adapted with permission.[130] Copyright © 2020 American Chemical Society.

**Figure 9.** Schematic representation of the promising characteristics of glyme-based electrolyte solutions for application in lithium-metal batteries using insertion (e.g., olivine-based compounds) as well as conversion cathodes (e.g., S-based and O$_2$-based electrodes); the typical relation between LiX salt concentration, Li$^+$ transport properties, safety, and stability of the glyme-based solutions is also shown.

| Chemical species | Acronym |
|---|---|
| nGlyme, n(ethylene glycol) dimethyl ether, $CH_3$-O-$(CH_2$-$CH_2$-O$)_n$-$CH_3$ | $G_n$ |
| Poly(ethylene glycol) dimethyl ether | PEGDME |
| Poly(ethylene oxide) | PEO |
| Poly(methyl methacrylate) | PMMA |
| Poly(butyl acrylate) | PBA |
| Polypyrrole | PPy |
| 1-3 Dioxolane | DOL |
| 1,2 Dimethoxy ethane | DME (or $G_1$) |
| Ethylene glycol diethyl ether | EG |
| Dimethylacetamide | DMA |
| Tetrahydrofuran | THF |
| 1,1,2,2-tetrafluoroethyl 2,2,3,3-tetrafluoropropyl ether, hydrofluoroether | HFE |
| Propylene carbonates | PC |
| Ethylene carbonates | EC |
| Dimethyl carbonates | DMC |
| Methylethylcarbonate, ethylmethylcarbonate | EMC |
| Diethyl carbonate | DEC |
| Fluoroethylene carbonate | FEC |
| Methylethylcarbonate | MEC |
| Acetonitrile | ACN |
| Dimethyl sulfoxide | DMSO |
| Dimethylformamide | DMF |
| Dimethylacetamide | DMA |
| Γ-butyrolactone | GBL |
| Tetraethylsulfamide | TESA |
| Lithium-bis-(trifluormethanesulfonyl)-imide, $LiN(SO_2CF_3)_2$, often LiTFSA | LiTFSI |
| Lithium bis(fluoro sulfonyl)imide, $LiN(SO_2F)_2$, often LiFSA | LiFSI |
| Cyclic imide, lithium 1,2,3-dithiazolidine-4,4,5,5-tetrafluoro-1,1,3,3-tetraoxide, $LiN(C_2F_4S_2O_4)$ | LiCTFSI |
| Lithium hexafluorophosphate | $LiPF_6$ |
| Lithium tetrafluoroborate | $LiBF_4$ |
| Lithium bis(perfluoroethanesulfonyl)imide, often LiBETA | LiBETI |
| Lithium tetrafluoroborate | LiTFB |
| Lithium trifluoromethylsulfonate, lithium triflate, $LiSO_3CF_3$ | LiTf |
| Lithium trifluoroacetate | LiTFA |
| Lithium 2-trifluoromethyl-4,5-dicyanoimidazole | LiTDA |
| Lithium tetra(trifluoromethanesulfonyl)propene | LiTFSP |
| Lithium bis(nonafluorobutanesulfonyl)imide, $LiN(C_4F_9SO_2)_2$ often LiNFSA | LiNFSI |
| Lithium nitrate | $LiNO_3$ |
| lithium bis(pentafluoroethylsulfonyl)imide, $LiN(SO_2C_2F_5)_2$ | LiBETI |
| Lithium bis(oxalato)borate, $LiC_4BO_8$ | LiBOB |
| Dilithium Dodecafluorododecaborate, $Li_2B_{12}F_{12}$ | $Li_2DFB$ |
| N, N-diethyl-N-methyI-N-(2-methoxyethyl) ammonium bis(trifluoromethanesulfonyl)imide | DEMETFSI |
| Methyl butyl pyrrolidinium-bis-(trifluoromethanesulfonyl)-imide | $Pyr_{14}TFSI$ |
| 1,1,2,2-Tetrafluoroethyl 2,2,2-trifluoroethyl ether | TFTFE |
| $LiNi_xMn_yCo_zO_2$ | NMC |
| Solid electrolyte interphase | SEI |
| Ionic liquid | IL |
| Solvate ionic liquid | SIL |
| Solvent separated ion pair | SSIP |
| Contact ion pair | CIP |
| Vinylene carbonate | VC |
| Vinylethylene carbonate | VEC |
| 1,3-Propane sultone | 13PS |
| Hard carbon | HC |
| Trimethyl phosphate | TMP |
| Triethyl phosphate | TEP |
| Tripropyl phosphate | TPrP |
| 2-(2,2,2-trifluoroethoxy)-1,3,2-dioxaphospholane 2-oxide | TFEP |
| 2,2,2-trifluoroethyl methyl carbonate | FEMC |
| Di-(2,2,2-trifluoroethyl)carbonate | DFDEC |
| 1,1,2,2-tetrafluoroethyl-2',2',2'-trifluoroethyl ether | TTFE |
| Ethoxy(pentafluoro)cyclotriphosphazene | PFPN |
| Polystyrene | PS |
| Poly(ethylene glycol) methacrylate | PEGMA |
| poly(ethylene glycol) dimethacrylate | PEGDMA |
| poly(ethylene glycol) methyl ether acrylate | PEGA |

**Table 1**

| Solvents | Safety hazards (SDS) | Melting point [°C] | Boiling point [°C] | Flash point [°C] | Density [g cm$^{-3}$] | Viscosity [mPa s] |
|---|---|---|---|---|---|---|
| DMC | - Flammable | 2 | 90 [27] | 18 [27] | 1.069 | 0.59 [28] |
| EC | - Elevated Toxicity<br>- Irritating | 35 | 248 [27] | 160 [27] | 1.321 | 2.56 [28] |
| PC | - Irritating | -55 | 242 [27] | 132 [27] | 1.204 | 2.5 [28] |
| EMC | - Flammable | -55 | 109 [27] | 23 [27] | 1.006 | 0.65 [29] |
| DEC | - Flammable | -43 | 126 [27] | 33 [27] | 0.975 | 0.753 [28] |
| DME | - Flammable<br>- Irritating<br>- May affect fertility | -58 | 84 [27] | -2 [27] | 0.867 | 0.42 – 0.46 [25] |
| DEGDME | - Flammable<br>- May affect fertility | -64 | 162 | 57 | 0.94 | 0.98 – 1.0 [25] |
| TREGDME | - Irritating,<br>- May affect fertility | -45 | 225 | 113 | 0.985 | 1.95 – 2.16 [25] |
| TEGDME | - May affect fertility | -30 | 276 [27] | 140 [27] | 1.009 | 3.3 – 3.7 [25] |
| PEGDME (250 g mol$^{-1}$) | / | -23 | 240 | 135 | 1.03 | 7.2 (20 °C) [30] |
| PEGDME (500 g mol$^{-1}$) | / | 13 | > 250 | 254 | 1.07 | 28 (20 °C) [30] |

**Table 2**

| Techniques and other acronyms | Acronym |
|---|---|
| Electrochemical quartz crystal microbalance | EQCM-A |
| Cyclic Voltammetry | CV |
| Electrochemical impedance spectroscopy | EIS |
| Very-low-frequency electrochemical impedance spectroscopy | VLF-EIS |
| Galvanostatic cycling | GC |
| Nuclear magnetic resonance | NMR |
| Pulse field gradient nuclear magnetic resonance | PFG-NMR |
| Pulsed-gradient spin-echo nuclear magnetic resonance | PGSE-NMR |
| Electrophoretic NMR | eNMR |
| First-principles molecular dynamics | FPMD |
| Electrochemical mass spectrometry | ECMS |
| Linear sweep voltammetry | LSV |
| Solvate ionic liquid | SIL |
| Contact ion pair | CIP |
| Scanning Electron Microscopy | SEM |
| Transmission Electron Microscopy | TEM |
| X-Ray Diffraction | XRD |
| Small-angle X-ray scattering | SAXS |
| In situ subtractively normalized Fourier transform infrared spectroscopy | SNIFTIRS |
| X-ray photoelectron spectroscopy | XPS |
| Thermogravimetric Analysis | TGA |
| Open circuit voltage | OCV |
| Gas diffusion layer | GDL |
| Oxygen reduction reaction | ORR |
| Oxygen evolution reaction | OER |
| Rotating ring disk electrode | RRDE |
| Mass spectrometry | MS |
| High-energy X-ray total scattering | HEXTS |
| Molecular dynamics | MD |
| Molecular weight | MW |
| Molecular orbital | MO |
| Density functional theory | DFT |
| Potentials of mean force | PMF |
| Infrared | IR |
| X-ray absorption spectroscopy | XAS |
| X-ray absorption near edge structure | XANES |
| Oxygen reduction reaction | ORR |
| Oxygen evolution reaction | OER |
| Donor number | DN |
| Highest occupied molecular orbital | HOMO |
| Lowest unoccupied molecular orbital | LUMO |

**Table 2**

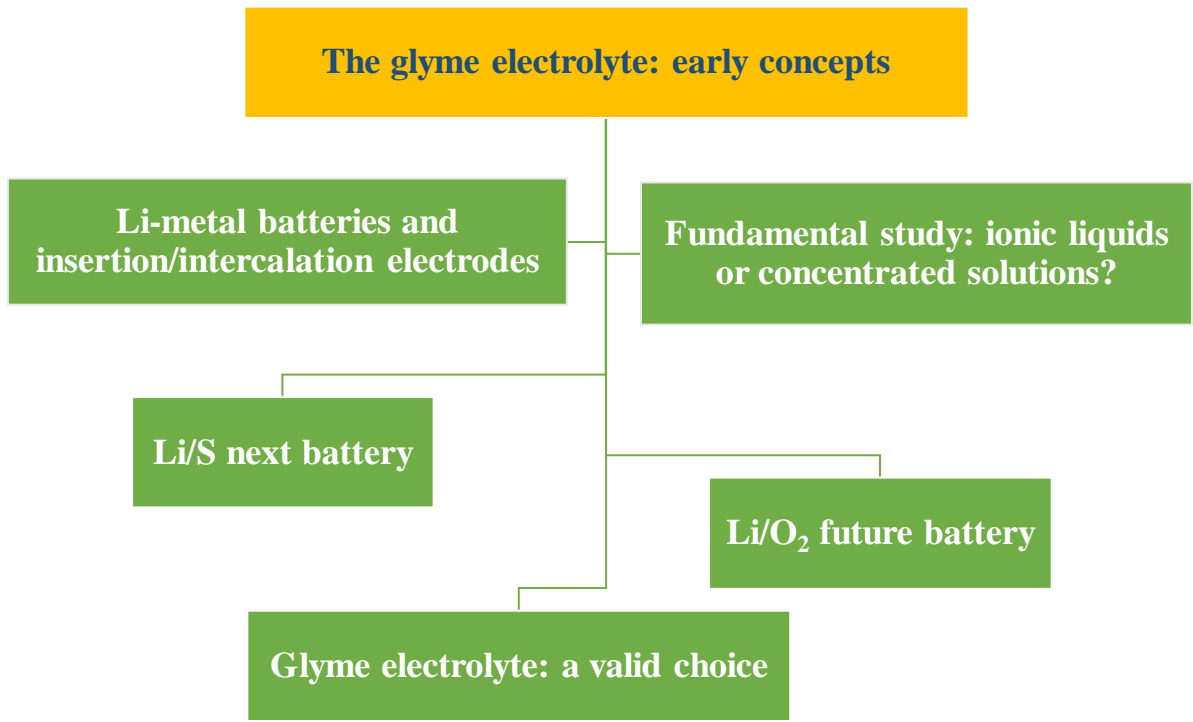
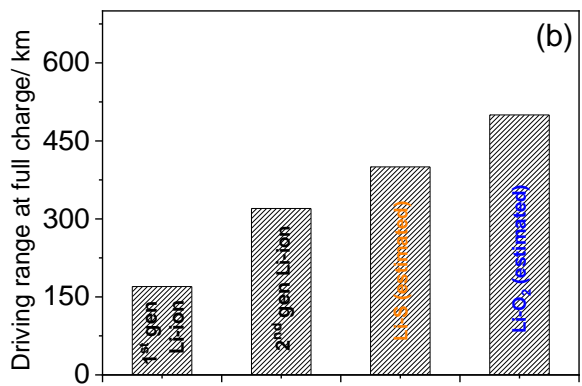
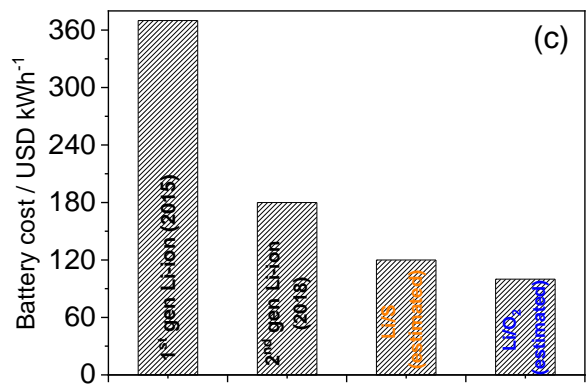

**Figure 1**

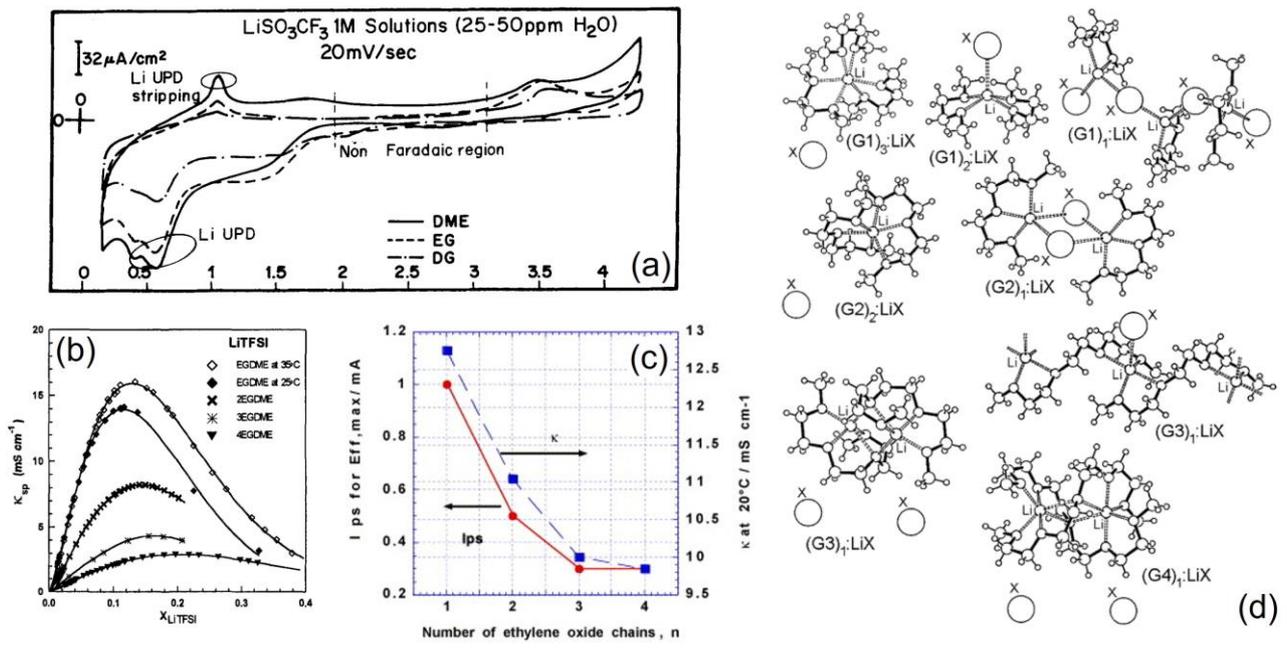

**Figure 2**

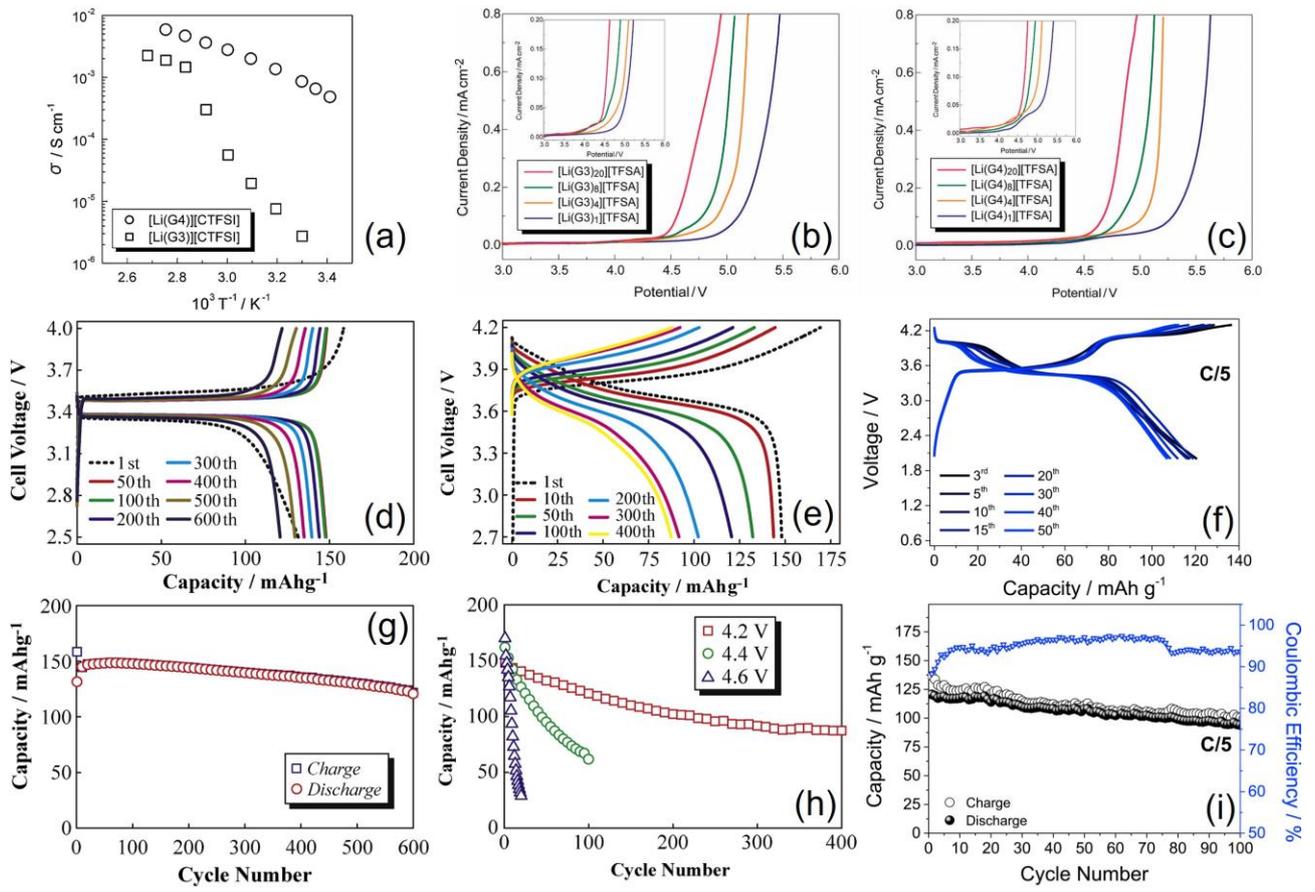

**Figure 3**

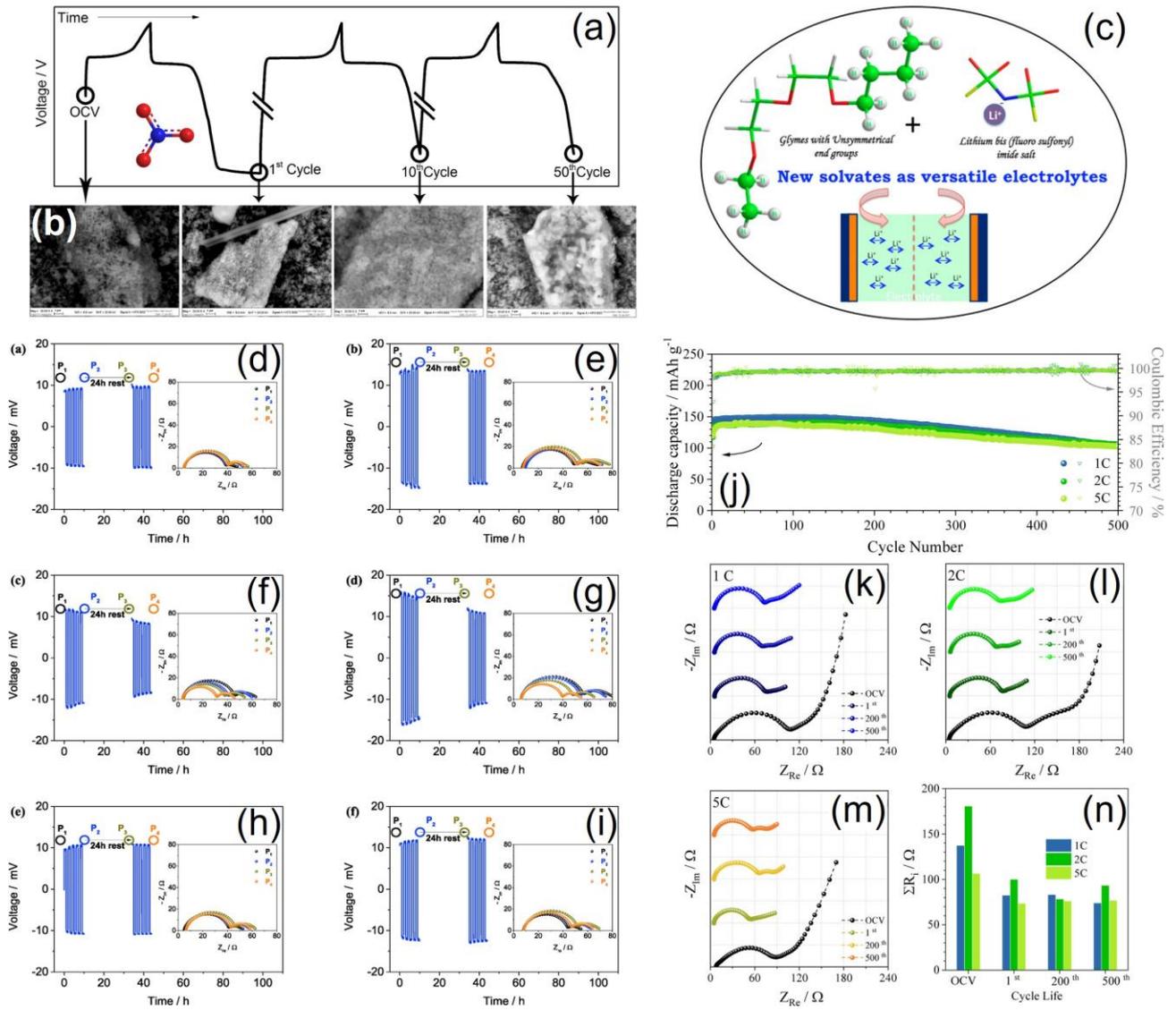

**Figure 4**

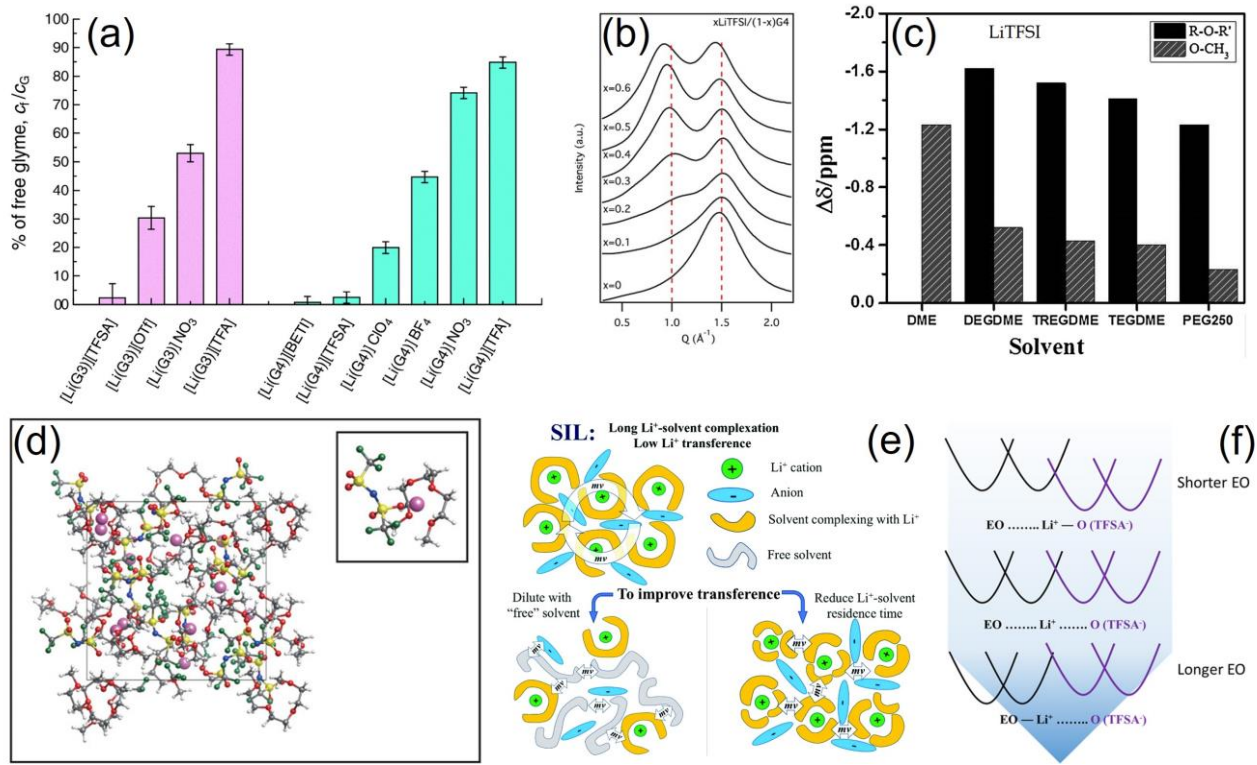

**Figure 5**

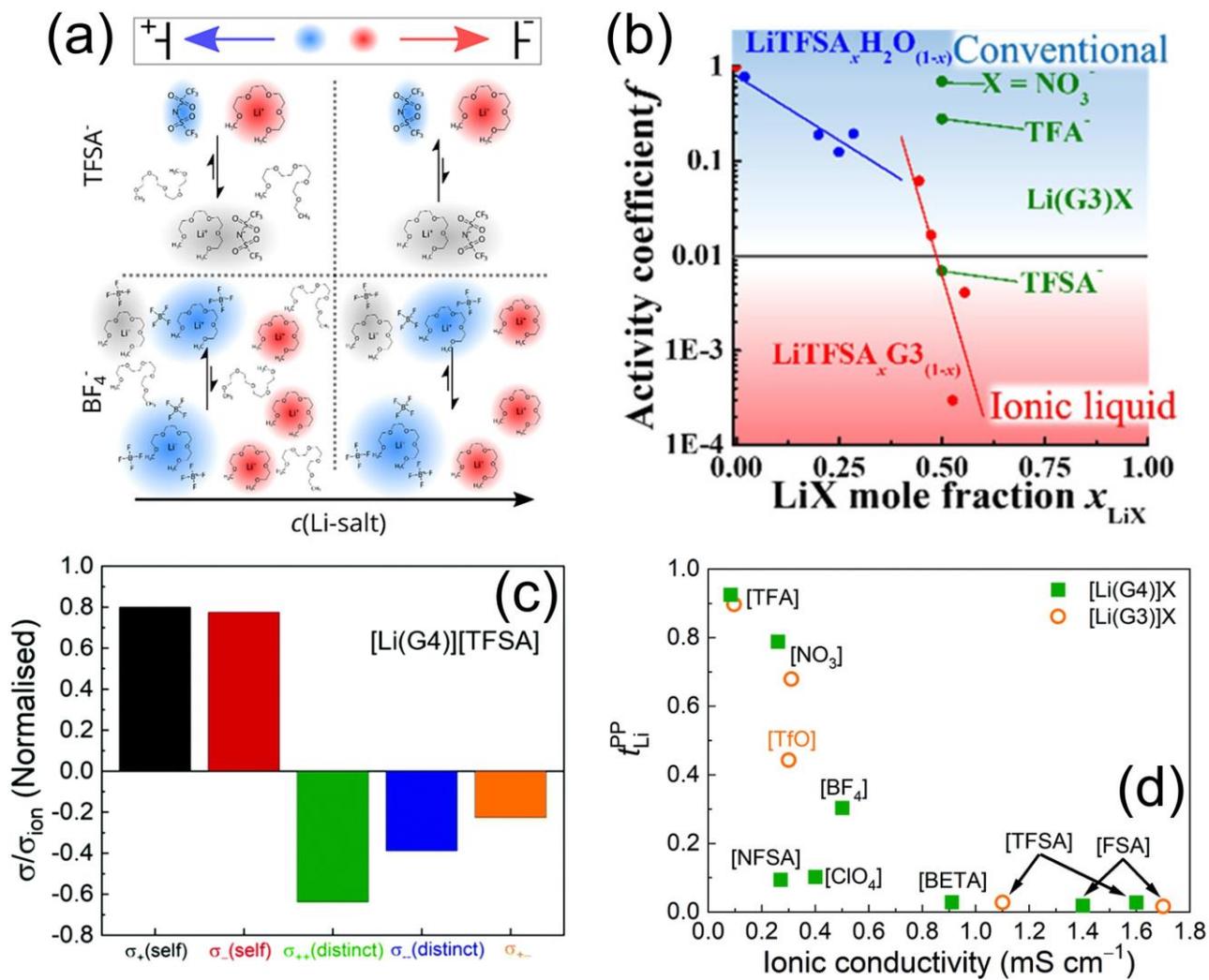

**Figure 6**

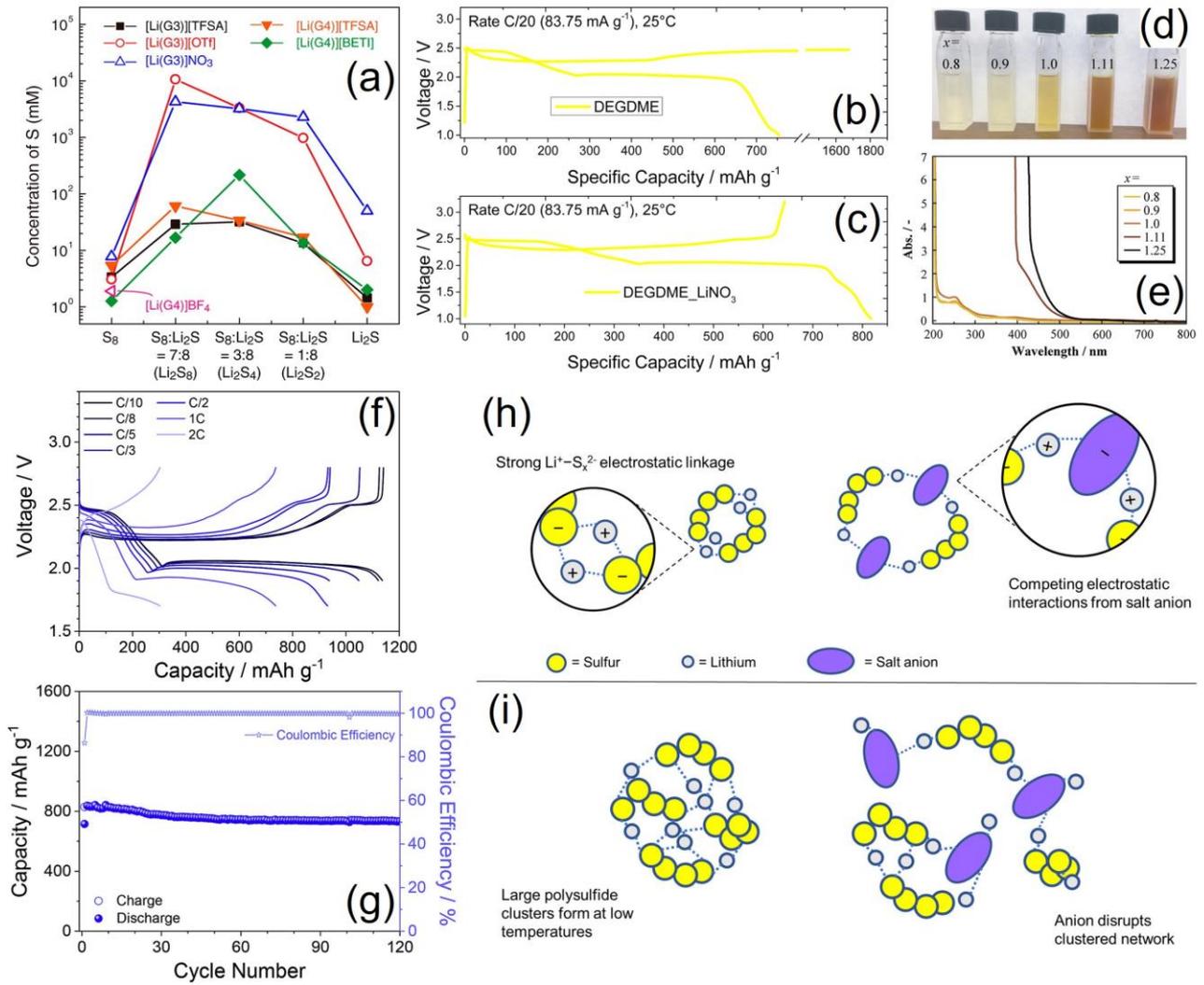

**Figure 7**

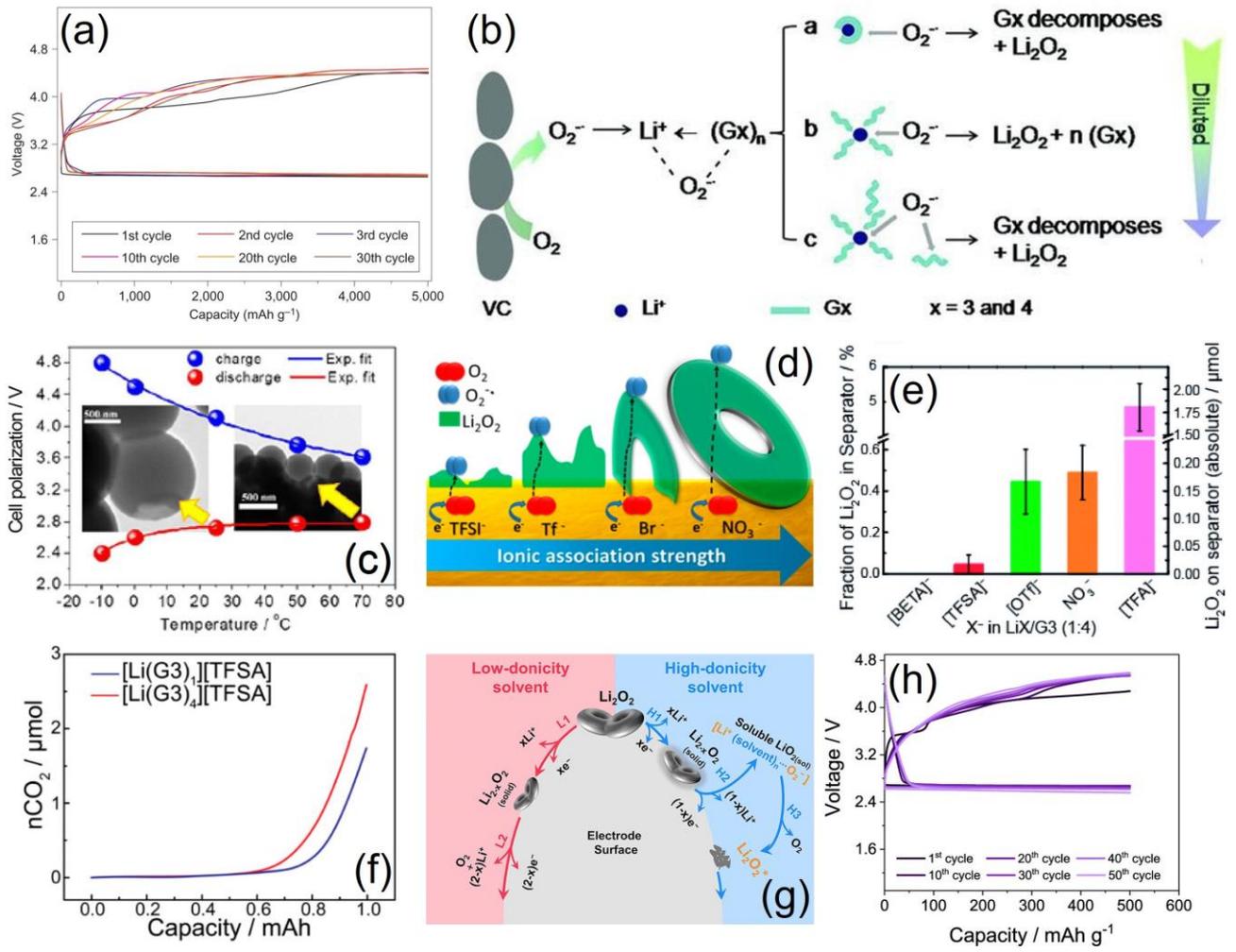

**Figure 8**

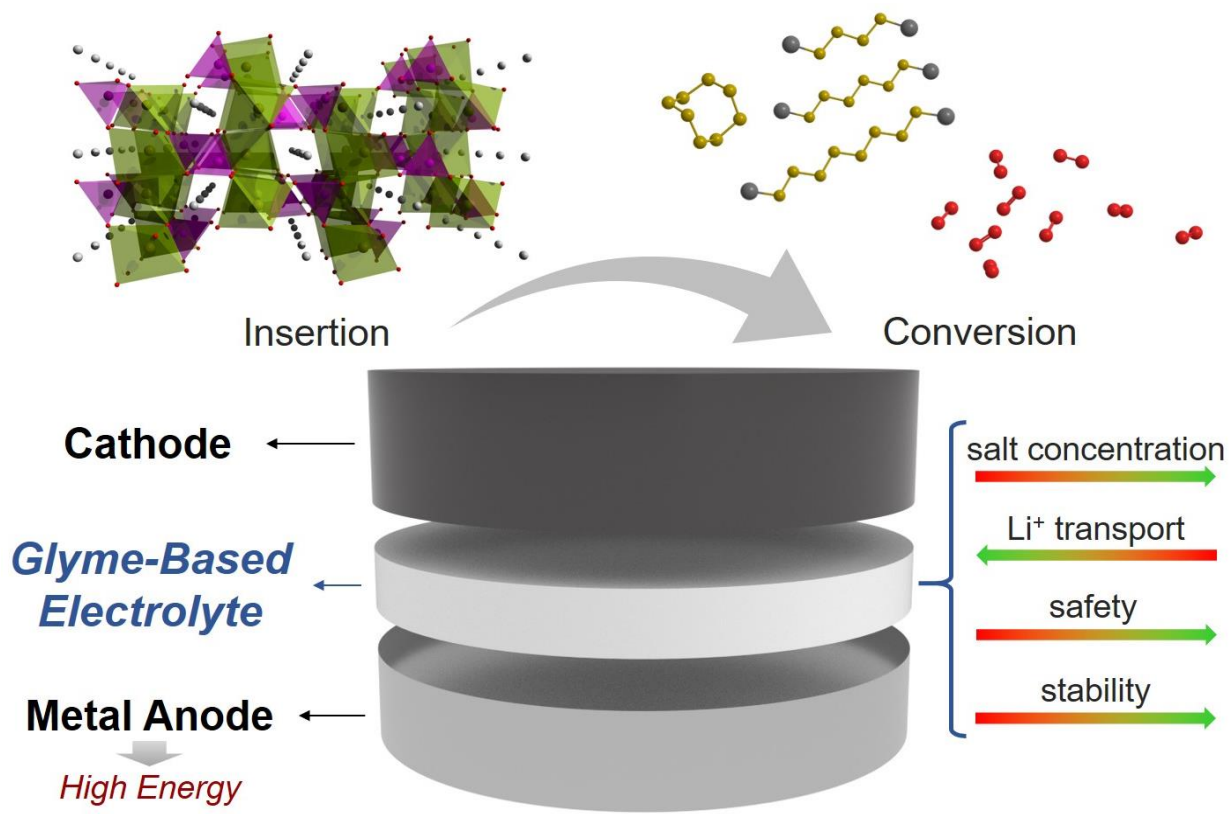

**Figure 9**